\documentclass[preprintnumbers,article,amsmath,amssymb,floatfix,10pt,prd,superscriptaddress,nofootinbib]{revtex4}

\usepackage{doi}
\usepackage{hyperref}
\hypersetup{
  colorlinks=true,        
  linkcolor=magenta,         
  citecolor=red,         
}

\providecommand{\dif}{\mathrm{d}} \def\d{\dif}
\usepackage{graphicx}
\usepackage{dcolumn}
\usepackage{bm}
\usepackage{color}
\usepackage{enumitem}
\usepackage{amsmath}
\usepackage{amssymb}
\def\EE{{\cal E}}
\def\LL{{\cal L}}
\newcommand{\beq}{\begin{equation}}
\newcommand{\eeq}{\end{equation}}
\newcommand{\bea}{\begin{eqnarray}}
\newcommand{\eea}{\end{eqnarray}}
\newcommand{\non}{\nonumber}
\usepackage{bbm}
\usepackage{amsfonts}
\usepackage{mathrsfs}
\usepackage{latexsym}
\usepackage{epsfig}
\usepackage{epstopdf}
\usepackage{epstopdf}
\usepackage{graphicx}
\usepackage{amssymb}
\usepackage{amsmath}
\usepackage{dcolumn}
\usepackage{bm}
\usepackage{color}
\usepackage{comment}
\usepackage{xcolor}
\usepackage{orcidlink}

\DeclareUnicodeCharacter{2217}{\ensuremath{\ast}}
\begin{document}
\title{ Relativistic Dynamics and Bondi-Hoyle-Lyttleton Accretion onto Rotating Embedded Black Hole Models}

\author{Asifa Ashraf}
\email{asifamustafa3828@gmail.com}\affiliation{School of
Mathematical Sciences, Zhejiang Normal University, Jinhua, Zhejiang
321004, China}

\author{Orhan Donmez}
\email{orhan.donmez@aum.edu.kw}
\affiliation{College of Engineering and Technology, American University of the Middle East, Egaila 54200, Kuwait}

\author{Abdelmalek Bouzenada}
\email{abdelmalekbouzenada@gmail.com}
\affiliation{Laboratory of Theoretical and Applied Physics, Echahid Cheikh Larbi Tebessi University 12001, Algeria}

\author{Chengxun~Yuan}
\email{yuancx@hit.edu.cn (Corresponding Author)}
\affiliation{School of Physics, Harbin Institute of Technology, Harbin 150001, People’s Republic of China}

\author{Aylin ÇALIŞKAN}
\email{aylincaliskan@istanbul.edu.tr}
\affiliation{Department of Physics, Faculty of Science,
Istanbul University, Istanbul 34134, Turkey}

\author{Gulzoda Rakhimova}
\email{rakhimovagulzoda96@gmail.com}
\affiliation{National Research University TIIAME, Kori Niyoziy 39, Tashkent 100000, Uzbekistan}
\affiliation{University of Tashkent for Applied Sciences, Str. Gavhar 1, Tashkent 100149, Uzbekistan}

\author{Ahmadjon Abdujabbarov}
\email{ahmadjon@astrin.uz}
\affiliation{School of Physics, Harbin Institute of Technology, Harbin 150001, People’s Republic of China}
\affiliation{National University of Uzbekistan, Tashkent 100174, Uzbekistan}
\affiliation{Tashkent State Technical University, Tashkent 100095, Uzbekistan}

\begin{abstract}

In this paper, we examine the motion of test particles and relativistic accretion mechanisms within the spacetime of a rotating and embedded black hole (BH). In this case, the geometric properties of the metric and their dynamical consequences for particle trajectories are systematically studied, with a specific focus on circular orbits together with their existence criteria and stability constraints. Also, the effective potential and the corresponding effective force are constructed to quantify the influence of rotation and embedding parameters on the attractive and repulsive sectors of the gravitational interaction. Closed-form expressions for orbital frequencies as measured by a distant observer are derived, enabling a quantitative analysis of relativistic precession phenomena, including periastron advance and Lense-Thirring precession. Furthermore, we conduct general-relativistic hydrodynamic simulations of Bondi-Hoyle-Lyttleton (BHL) accretion onto rotating embedded BHs. In addition, within the framework of the BHL accretion mechanism, the numerical solution of the GRH equations shows that the embedding parameter $\alpha$ systematically modifies the morphology of the shock cone formed around embedded BHs compared to the Kerr model. In particular, a wider opening angle of the cone is produced, the compression of matter in the post-shock region is weakened, and the dynamical variability of the flow is enhanced. The time-dependent mass accretion rate exhibits increasing oscillation amplitudes and long-term variations with increasing $\alpha$, while these amplitudes are found to be suppressed by the frame-dragging effect associated with the BH spin parameter. At the same time, increasing values of $\alpha$ lead to a strengthening of the QPO frequencies formed around embedded BHs in the LFQPO regime, enhancing their observability and increasing the likelihood of detecting commensurate frequency ratios such as $3{:}2$.
\end{abstract}

\maketitle

\date{\today}


\section{Introduction}\label{S1}

General relativity (GR) predicts the formation of black holes (BHs), and this prediction has received strong empirical support from recent advances in strong-field gravity, notably through the direct observation of gravitational waves emitted during compact binary coalescences \cite{BHP1, BHP2, BHP3} and the imaging of supermassive BH shadows by the Event Horizon Telescope \cite{BHP4, BHP5, BHP6, BHP7, BHP8}. In this case, these observations probe the near-horizon regime and provide direct tests of GR in highly nonlinear gravitational fields. Nevertheless, a variety of compact astrophysical configurations, such as wormholes and some classes of quasars, can generate observational features that closely resemble those associated with BHs and are therefore commonly referred to as BH mimickers. At the same time, GR is known to be incomplete at both theoretical and observational levels, facing persistent issues including the explanation of the cosmic expansion history, the emergence of large-scale structure, the treatment of spacetime singularities, and the formulation of a consistent quantum description of gravity. Also, in this context, limitations and uncertainties in current observational data allow for the possibility that gravity may deviate from GR, thereby motivating the study of extended or modified gravitational frameworks with additional degrees of freedom and richer dynamics. A central aspect of these investigations is the singularity problem, which is generally expected to be resolved by quantum gravity effects. However, an early classical resolution was proposed by Bardeen \cite{BHP9}, who introduced a static regular BH characterized by a nonsingular core. Later, this solution was shown to emerge from Einstein gravity coupled to a nonlinear magnetic monopole field \cite{BHP10}, providing a concrete physical realization. Following this result, a wide class of regular (nonsingular) BH solutions has been developed, which can be classified into two main categories: models obtained by solving modified field equations with appropriate matter sources in generalized gravity theories, often displaying effective semiclassical behavior \cite{BHP11, BHP12, BHP13, BHP14}, and constructions interpreted as quantum-corrected extensions of classical singular BH geometries \cite{BHP15, BHP16, BHP17, BHP18, BHP19}. In the latter case, quantum gravitational corrections act to regularize the spacetime curvature, illustrtaing that the formation of singularities may be avoided. Consequently, regular BHs constitute a useful theoretical laboratory for investigating the classical limit of quantum BHs in the absence of a complete theory of quantum gravity, with detailed reviews and systematic discussions presented in Refs. \cite{BHP20,BHP21}. 

Quasi-periodic oscillations (QPOs)  \cite{QPOs1, QPOs2, QPOs3, QPOs4} constitute one of the most reliable timing phenomena detected in high-energy astrophysical sources and are manifested as narrow and coherent features in the X-ray power density spectra of accreting neutron stars and BH binaries \cite{QPOs5, QPOs6}. In this context, the wide interval of observed QPOs frequencies serves as a direct diagnostic of the dynamical state of matter in the innermost regions of accretion disks, where relativistic gravity governs plasma motion in the vicinity of compact objects \cite{QPOs7, QPOs8,Donmez:2013qxa,Donmez:2017gdp}. In general, low-frequency QPOs are associated with large-scale disk dynamics, such as global disk oscillations, Lense-Thirring precession, or magnetohydrodynamic processes, whereas high-frequency QPOs, typically detected in the $10^2$-$10^3$ Hz range, are linked to oscillatory mechanisms controlled by orbital motion and epicyclic frequencies in strong gravitational fields \cite{QPOs9, QPOs10, QPOs11, QPOs12, QPOs13, QPOs14, QPOs15, QPOs16}. Observational studies of microquasars, including $GRO J1655$-$40$ \cite{QPOs17}, $XTE J1550$-$564$ \cite{QPOs18}, $GRS 1915$+105 \cite{QPOs19}, and $H1743$-$322$, consistently reveal twin high-frequency peaks exhibiting a characteristic $3{:}2$ ratio \cite{QPOs20}. This behavior is commonly interpreted as the manifestation of a non-linear resonance between radial and vertical epicyclic modes of particle motion in the disk \cite{QPOs21}. Such an interpretation naturally follows from GR, where the separation of fundamental frequencies near compact objects allows for resonant coupling between oscillatory modes \cite{QPOs22, QPOs23, QPOs24}. Moreover, it provides a quantitative explanation for the empirical scaling relation $\nu \propto 1/M$, enabling QPO measurements to constrain BH mass and spin parameters \cite{QPOs25}, frequently indicating rapidly rotating BHs with spin values in the range $a \approx 0.8$-$0.96$ \cite{QPOs26}. Although alternative mechanisms, including diskoseismic oscillations, orbiting hot spots, and higher-order harmonic structures, have been proposed \cite{QPOs27, QPOs28}, the resonance framework remains consistent with the majority of high-frequency observational data. Comparable oscillatory signals have also been reported in neutron star and white dwarf systems,  however, stable high-frequency twin resonances are predominantly observed in BH sources, supporting their interpretation as probes of strong-field gravity \cite{QPOs29, QPOs30}. Consequently, QPOs have become an effective observational tool for investigating spacetime geometry in the immediate vicinity of compact objects \cite{QPOs31, QPOs32}, and forthcoming X-ray missions are expected to extend these studies to ULXs and AGNs \cite{QPOs33, QPOs34, QPOs35}. In parallel, recent theoretical studies have tested QPO models to include modified gravity and quantum-corrected (QC) BH spacetimes \cite{RefQS1, RefQS2, RefQS3, RefQS4, RefQS5, RefQS6, RefQS7, RefQS8}, in which deviations from classical GR can lead to observable modifications of orbital dynamics. In particular, braneworld models and rotating braneworld Kerr geometries \cite{RefQS9, RefQS10, RefQS11, RefQS12, RefQS13, RefQS14, RefQS15} introduce corrections to particle motion and induce shifts in the ISCO's radius \cite{RefQS16, RefQS17, RefQS18, RefQS19}, resulting in illustrated QPO spectra that may be detectable with high-precision timing observations. QC BH models further modify this picture, as GUP-induced effects change the ISCO's location and directly influence high-frequency QPO characteristics \cite{RefQS20, RefQS21, RefQS22}, while additional deviations arise within loop quantum gravity approaches \cite{RefQS23, RefQS24, RefQS25, RefQS26} and non-commutative geometry frameworks \cite{RefQS27}. Also, these results demonstrate that QPOs represent a sensitive observational channel for testing strong-field GR and for placing constraints on possible quantum-gravity corrections in astrophysical BH systems \cite{RefQS28, RefQS29}.

This paper is organized as follows. In Sec. (\ref{S2}), we present the rotating and embedded BH metric model and describe its fundamental geometrical and physical characteristics relevant for particle dynamics. Subsection (\ref{S2-1}) is illustrated by a comprehensive analysis of circular orbits around the rotating and embedded BH, where the criteria for their existence and stability are systematically investigated. Also, in Subsec. (\ref{S2-2}), we examine the effective potential associated with test particle motion in the considered spacetime and explain how rotation and embedding parameters alter its structure. Subsec. (\ref{S2-3}) concentrates on the corresponding effective force, providing a quantitative description of attractive and repulsive regimes generated by the combined influence of rotation and embedding. Also, the exact orbital frequencies measured by a distant observer are obtained in Subsec. (\ref{S3-1}), followed in Subsec. (\ref{S3-2}) by a detailed analysis of periastron advance and Lense-Thirring precession in the rotating embedded BH background. In Sec. (\ref{S4}), we report general-relativistic hydrodynamic simulations of BHL accretion onto rotating embedded BHs, shows the dynamical interaction between the spacetime geometry and accretion flows. Also, subsections (\ref{S4_1}) and (\ref{S4_2}) investigate, respectively, the morphological transitions of the shock cone with angular density modulation and the time-dependent accretion dynamics together with their variability signatures. In Subsec. (\ref{S4_3}), we study QPOs imprints produced by embedding effects in BHL accretion flows and discuss their possible observability. In this context, Sec. (\ref{S5}) explains the principal results and indicates potential directions for future investigations.

\section{Rotating and embedded Black Hole Metric Model}\label{S2}
In this part, the geometry describing a rotating BH embedded within an external geometric background is illustrated by the following spacetime line element \cite{metric}:
\begin{equation}\label{BH}
ds^{2}=-\left(1-\frac{2r\,f(r)}{\rho^{2}}\right)dt+\left(\frac{\rho^2}{\Delta}\right)dr^{2}+\rho^{2}d\theta^{2}-\left(\frac{4arf(r)\sin^{2}\theta}{\rho^{2}}\right)dtd\phi+\left(\frac{\Sigma\,\sin^{2}\theta}{\rho^{2}}\right)d\phi^{2},
\end{equation}
where the radial and angular dependence of the metric is encoded in the functions 
\bea
\Delta &=&  a^2 - 2 r f(r) +r^2, \quad f(r) = M - \frac{4 \alpha }{r},\\\
\rho^2 &=& r^2 + a^2\, \cos^2\theta,\\\
\Sigma &=& \left(a^2+r^2\right)^2-a^2 \, \Delta.
\eea
In this formulation, the parameter $M$ represents the gravitational mass of the BH model, while $a$ explains the influence of specific angular momentum associated with rotation. Also, the quantity $\alpha$ parametrizes corrections induced by extrinsic curvature effects, which modify the effective mass function $f(r)$ and, in this context, introduce deviations from the standard vacuum solution. Consequently, the spacetime structure depends explicitly on both rotational effects and embedding-induced corrections. In the limiting case $\alpha=0$, the function $f(r)$ reduces to a constant mass term, and the metric (Eq. \ref{BH}) exactly recovers the Kerr geometry. Moreover, by further setting $a=0$ together with $\alpha=0$, the line element consistently reduces to the Schwarzschild BH spacetime, confirming the internal consistency of the model and its correspondence with well-known solutions.

\subsection{Testing Circular orbits information around rotating and embedded BH model}\label{S2-1} 

\begin{figure*}
\centering 
\includegraphics[width=\hsize]{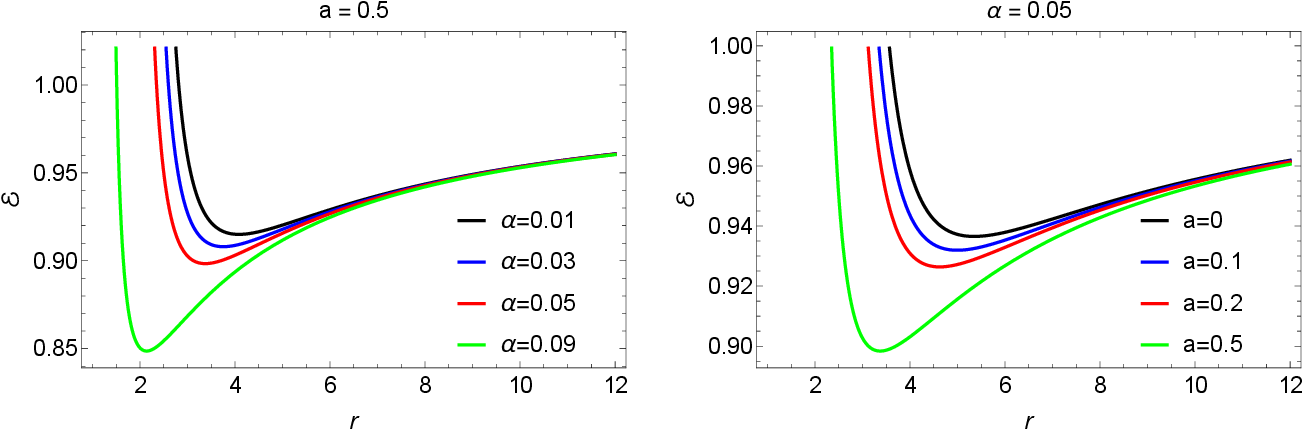}
\caption{Radial behavior of the conserved energy of neutral test particles moving on circular orbits around a rotating and embedded BH. The profiles illustrate the dependence of the particle energy on the spacetime parameters and orbital radius\label{F1}.}
\label{figENG}
\end{figure*}

\begin{figure*}
\centering 
\includegraphics[width=\hsize]{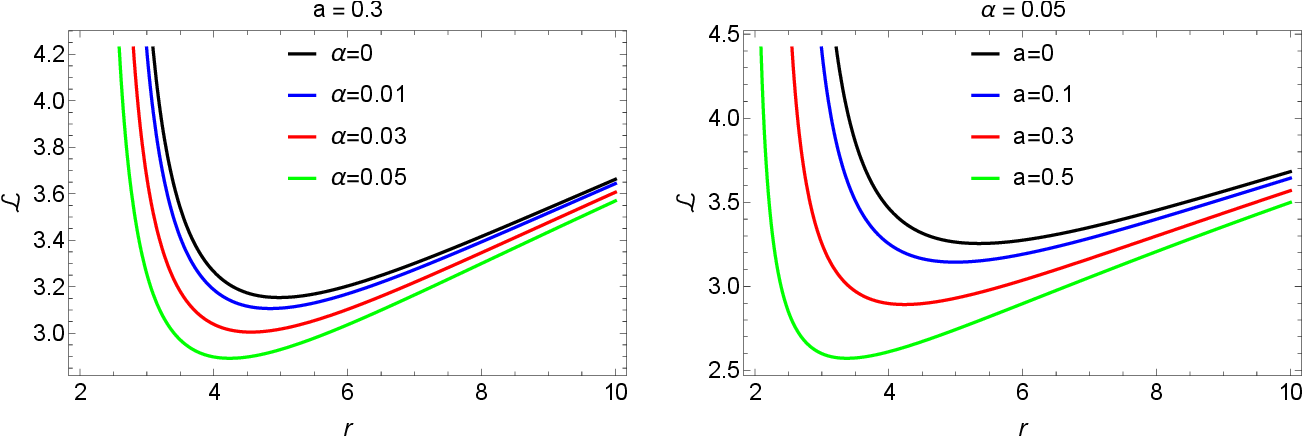}
\caption{Radial distribution of the conserved angular momentum of neutral test particles orbiting a rotating and embedded BH, showing how the spacetime rotation and embedding parameters modify the angular momentum requirement for stable motion\label{F2}.}
\label{fig_ANG}
\end{figure*}

In this part, the motion of a neutral test particle in the spacetime under consideration is described using the Hamiltonian formalism. In this approach, the particle dynamics follow from the Hamiltonian function \cite{B1}-\cite{B12}:
\beq\label{Ham}
H=\frac{1}{2}g^{\alpha \beta} p_{\alpha} p_{\beta} + \frac{1}{2}m^2,
\eeq
where $m$ defines the rest mass of the particle. Also, the four-momentum is defined as $p^{\gamma}=m u^{\gamma}$, with the four-velocity given by $u^{\gamma}=dx^{\gamma}/d\tau$, and $\tau$ illustrating the proper time measured along the particle trajectory. The evolution of the phase-space variables is governed by the Hamilton equations \cite{B1}-\cite{B12}:
\beq
\frac{dx^{\gamma}}{d\zeta}\equiv m u^{\gamma}=\frac{\partial H}{\partial p_{\gamma}}, \quad
\frac{d p_{\gamma}}{d\zeta} = -\frac{\partial H}{\partial x^{\gamma}},
\eeq
where the affine parameter is introduced as $\zeta=\tau/m$, which simplifies the formal structure of the equations of motion.

Because the BH spacetime is stationary and axisymmetric, the system admits two independent constants of motion associated with the Killing vectors $\partial_t$ and $\partial_\phi$. Also, these conserved quantities correspond to the specific energy $E$ and the specific angular momentum $L$ of the test particle and are expressed as \cite{B1}-\cite{B12}:
\bea\label{EE}
\frac{p_{t}}{m}&=&g_{tt}u^{t}+g_{t\phi}u^{\phi}=-\mathcal{E}, \\\label{LL}
\frac{p_{\phi}}{m}&=&g_{\phi\phi}u^{\phi}+g_{t\phi}u^{t}=\mathcal{L},
\eea
where $\mathcal{E}=E/m$ and $\mathcal{L}=L/m$ are the dimensionless forms of the energy and angular momentum, respectively. These quantities play a central role in characterizing the orbital structure and stability properties of particle motion.

Restricting the analysis to equatorial trajectories in the rotating and embedded BH background, the Hamiltonian in Eq. (\ref{Ham}) can be written explicitly as
\begin{equation}
H=   \frac{1}{2r^{2}\left(a^{2}+8\alpha+r^{2}-2r\right)}\left[ \mathcal{H}_{1}+\mathcal{H}_{2}+\mathcal{H}_{3}\right]
\end{equation}
where the terms $\mathcal{H}_{1}$, $\mathcal{H}_{2}$, and $\mathcal{H}_{3}$ collect contributions arising from the rotation parameter, the embedding parameter, and the radial and angular momenta. These terms are given by
\begin{equation}
\mathcal{H}_{1}=a^{2}\left(-\left(\EE^{2}\left(-8\alpha+r^{2}+2r\right)\right)+p_{\theta}^{2}+16\alpha p_{r}^{2}+2p_{r}^{2}r^{2}-4p_{r}^{2}r+r^{2}\right)    
\end{equation}
\begin{equation}
\mathcal{H}_{2}= 4a\EE\LL(r-4\alpha)-\EE^{2}r^{4}+\LL^{2}\left(8\alpha+r^{2}-2r\right)+8\alpha p_{\theta}^{2}+p_{\theta}^{2}r^{2}-2p_{\theta}^{2}r   
\end{equation}
and
\begin{equation}
\mathcal{H}_{3}= 64\alpha^{2}p_{r}^{2}+p_{r}^{2}r^{4}+a^{4}p_{r}^{2}-4p_{r}^{2}r^{3}+16\alpha p_{r}^{2}r^{2}+4p_{r}^{2}r^{2}-32\alpha p_{r}^{2}r+r^{4}-2r^{3}+8\alpha r^{2}   
\end{equation}
In this formulation, the Hamilton equations yield the full set of coupled differential equations governing the radial, angular, and temporal evolution of the test particle. This framework allows a systematic investigation of orbital properties, energy conditions, and angular momentum distributions in the rotating and embedded BH spacetime.
\bea
\frac{\mathrm{d} r}{\mathrm{d} \tau} &=& \frac{p_r \left(a^2+8 \alpha +r^2-2 r\right)}{r^2},\\\
\frac{\mathrm{d} \theta}{\mathrm{d} \tau} &=& \frac{p_\theta}{r^2},\\\
\frac{\mathrm{d} p_\theta}{\mathrm{d} \tau} &=& 0,\\\
\frac{\mathrm{d} \phi}{\mathrm{d} \tau} &=& \frac{2 a \EE (r-4 \alpha )+\LL \left(8 \alpha +r^2-2 r\right)}{r^2 \left(a^2+8 \alpha +r^2-2 r\right)},\\\non
        \frac{\mathrm{d} p_r}{\mathrm{d} \tau} &=& \frac{1}{r^{3}\left(a^{2}+8\alpha+r^{2}-2r\right)^{2}}\left[\mathcal{P}_{1}+\mathcal{P}_{2}+\mathcal{P}_{3}+\mathcal{P}_{4}+\mathcal{P}_{5}+\mathcal{P}_{6}\right],
\eea
where
\begin{equation}
\mathcal{P}_{1}= a^{6}p_{r}^{2}+a^{4}\left(p_{r}^{2}\left(24\alpha+2r^{2}-5r\right)-\EE^{2}(r-8\alpha)\right)   
\end{equation}
\begin{equation}
\mathcal{P}_{2}=2a^{3}\EE\LL(r-8\alpha)\non+a^{2}(-2\EE^{2}\left(-32\alpha^{2}+r^{3}-2(4\alpha+1)r^{2}+16\alpha r\right)    
\end{equation}
\begin{equation}
\mathcal{P}_{3}= -\LL^{2}(r-8\alpha)+p_{r}^{2}(192\alpha^{2}+r^{4}-6r^{3}\non+8(4\alpha+1)r^{2}-80\alpha r))+p_{\theta}(\tau)^{2}\left(a^{2}+8\alpha+r^{2}-2r\right)^{2}   
\end{equation}
\begin{equation}
\mathcal{P}_{4}=2a\EE\LL(-64\alpha^{2}+3r^{3}-4(4\alpha+1)r^{2}\non+32\alpha r)-\EE^{2}r^{5}+8\alpha\EE^{2}r^{4}+64\alpha^{2}\LL^{2}+\LL^{2}r^{4}    
\end{equation}
\begin{equation}
\mathcal{P}_{5}= -4\LL^{2}r^{3}+16\alpha\LL^{2}r^{2}+4\LL^{2}r^{2}-32\alpha\LL^{2}r+512\alpha^{3}p_{r}^{2}-p_{r}^{2}r^{5}+8\alpha p_{r}^{2}r^{4}   
\end{equation}
and 
\begin{equation}
\mathcal{P}_{6}=4p_{r}^{2}r^{4}-48\alpha p_{r}^{2}r^{3}-4p_{r}^{2}r^{3}+128\alpha^{2}p_{r}^{2}r^{2}+64\alpha p_{r}^{2}r^{2}-320\alpha^{2}p_{r}^{2}r    
\end{equation}
The constants of motion governing timelike geodesics, namely the specific energy $\mathcal{E}$ and the specific angular momentum $\mathcal{L}$ of test particles confined to the equatorial plane of a rotating and embedded BH spacetime, follow directly from the stationarity and axial symmetry of the geometry. For circular motion, these conserved quantities can be written as
\begin{eqnarray}
\label{r19}
\mathcal{E}&&= \left(\frac{-a^{2}(r-8\alpha)+2a(r-4\alpha)\sqrt{r-8\alpha}+r^{2}\left(8\alpha+r^{2}-2r\right)}{\left(r^{5}-a^{2}r(r-8\alpha)\right)\mathcal{D}}\right),\\\label{r20}
\mathcal{L}&&= \left(\frac{-a^{3}(r-8\alpha)+a^{2}\left(-8\alpha+r^{2}+2r\right)\sqrt{r-8\alpha}+ar^{2}(16\alpha-3r)+r^{4}\sqrt{r-8\alpha}}{\left(r^{5}-a^{2}r(r-8\alpha)\right)\mathcal{D}}\right),
\end{eqnarray}
where the factor $\mathcal{D}$ ensures proper normalization of the four-velocity and encodes the combined contribution of rotation and embedding effects, is defined as
\begin{equation}
\mathcal{D}=\left(\frac{2a^{3}(r-8\alpha)^{3/2}+a^{2}\left(-128\alpha^{2}-3r^{3}+3(8\alpha-1)r^{2}+40\alpha r\right)+2ar^{2}(3r-16\alpha)\sqrt{r-8\alpha}+r^{4}\left(16\alpha+r^{2}-3r\right)}{\left(r^{4}-a^{2}(r-8\alpha)\right)^{2}}\right)^{1/2}    
\end{equation}
In this case, $\mathcal{E}$ and $\mathcal{L}$ depend explicitly on the radial coordinate $r$, the rotation parameter $a$, and the embedding parameter $\alpha$, illustrating the relation between centrifugal, gravitational, and frame-dragging effects in the spacetime.

\begin{itemize}
    \item Figure (\ref{figENG}) displays the radial profiles of the specific energy for circular orbits. In these results, the first column isolates the role of the embedding parameter $\alpha$, while the second column illustrates the influence of the rotation parameter $a$. In this context, both parameters modify the depth and slope of the energy curves in a similar manner. Also, for fixed $r$, lower values of $a$ and $\alpha$ correspond to higher values of $\mathcal{E}$, indicating less efficient binding. Conversely, increasing either $a$ or $\alpha$ lowers the particle energy, which implies stronger relativistic effects and enhanced gravitational binding near the BH.

    \item The corresponding behavior of the specific angular momentum is shown in Fig. (\ref{fig_ANG}). The dependence on $\alpha$ is presented in the first column, whereas the second column emphasizes the variation with the rotation parameter $a$. In this case, the two parameters again produce comparable qualitative effects. For weak rotation and small embedding contributions, the angular momentum required to sustain circular motion is larger. As either $a$ or $\alpha$ increases, $\mathcal{L}$ decreases, indicating that frame dragging and embedding effects partially support orbital motion. Also, when the BH parameters are held fixed, the angular momentum increased with the radial coordinate $r$, consistent with the expected influence of circular geodesics in rotating spacetimes.
\end{itemize}

\begin{figure*}
\centering 
\includegraphics[width=\hsize]{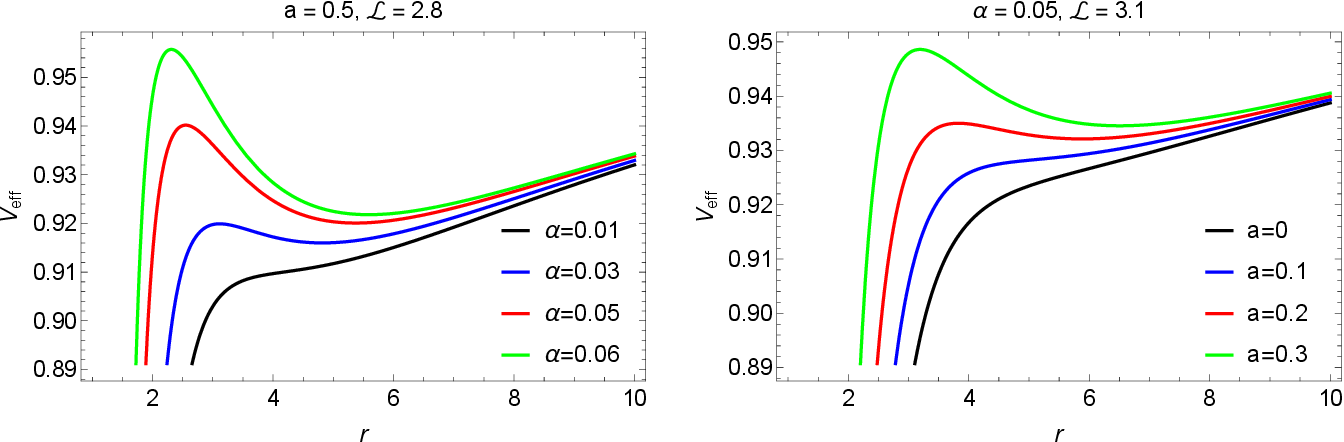}
\caption{Testing Effective potential for particles around a rotating and embedded BH\label{F3}.}\label{fig_eff}
\end{figure*}

\subsection{Testing the effective potential under rotating embedded BH model}\label{S2-2}

In this part, the motion of a massive test particle is constrained by the normalization of the four-velocity, $g_{\nu\sigma}u^{\nu}u^{\sigma}=-1$. Also, using this condition, the radial and polar contributions to the dynamics can be isolated and written in terms of an effective potential as \cite{B1}-\cite{B12}:
\beq
V_{eff}(r, \theta) = g_{rr}\, \Dot{r}^{2} + g_{\theta \theta}\, \Dot{\theta}^{2},
\eeq
where $\Dot{r}=dr/d\tau$ and $\Dot{\theta}=d\theta/d\tau$ denote derivatives with respect to the proper time $\tau$. In this formulation, the effective potential $V_{\rm eff}$ encodes the influence of the spacetime geometry on the particle motion through the conserved quantities associated with the Killing vectors. It can be expressed explicitly as \cite{B1}-\cite{B12}
\beq
V_{eff}(r, \theta) = \frac{\EE^2 g_{\phi \phi} + 2 \EE \LL g_{t \phi} + \LL^2 g_{tt}}{g_{t\phi}^{2} - g_{tt} g_{\phi\phi}} - 1.
\eeq
where $\EE$ and $\LL$ represent the conserved energy and angular momentum per unit mass, respectively. Also, this expression makes clear that frame-dragging effects, encoded in the off-diagonal metric component $g_{t\phi}$, directly contribute to the structure of the effective potential.

For the spacetime under consideration, the effective potential simplifies to a purely radial function and takes the form
\bea\non
V_{\rm eff}(r) &=& \frac{a^2 r^2 \chi +r^2 \left(8 \alpha +r^2-2 r\right)  \chi + 2 a \LL (r-4 \alpha )}{a^2 \left(-8 \alpha +r^2 + 2 r\right)+r^4},
\eea
where the auxiliary function $\chi$ is defined as
\beq
\chi = \sqrt{\frac{a^2 \left(-8 \alpha +r^2+2 r\right)+r^2 \left(\LL^2+r^2\right)}{r^2 \left(a^2+8 \alpha +r^2-2 r\right)}}
\eeq
The function $\chi$ arises from the normalization condition and ensures the consistency of the effective potential with the conserved quantities. Its explicit dependence on $a$, $\alpha$, and $\LL$ reflects the combined influence of rotation and spacetime parameters on the particle dynamics.

The effective potential $V_{\rm eff}(r,\theta)$ provides a compact and physically transparent framework to investigate the qualitative behavior of test particle orbits without solving the full set of geodesic equations. In particular, circular motion confined to the equatorial plane, $\theta=\pi/2$, is determined by imposing the standard conditions \cite{B1}-\cite{B12}:
\beq
V_{\rm eff}(r) = 0, \quad \frac{\d V_{\rm eff} (r)}{\d r} = 0.\label{Veff-1}
\eeq
The first condition ensures that the radial velocity vanishes, while the second guarantees radial force balance.

Extrema of the effective potential correspond to circular orbits. A minimum of $V_{\rm eff}$ indicates a stable circular orbit, whereas a maximum corresponds to an unstable configuration. In the Newtonian limit, the presence of a minimum defines the smallest stable circular orbit for a given angular momentum. In relativistic spacetimes, however, rotation and additional geometric parameters modify the depth and location of the extrema of $V_{\rm eff}$, leading to shifts in the orbital structure. For the Schwarzschild BH in GR, two extrema are present for fixed angular momentum. In the present model, the location of the ISCOs is primarily controlled by the value of the angular momentum determined from Eq. (\ref{Veff-1}). Fig. (\ref{fig_eff}) illustrates the radial behavior of $V_{\rm eff}$ for different values of the black hole parameters. The first column emphasizes the effect of the parameter $\alpha$, while the second column shows the influence of the rotation parameter $a$. Both parameters exhibit a similar qualitative behavior: increasing either $\alpha$ or $a$ reduces the depth of the potential well and enhances orbital instability. Consequently, stable circular orbits gradually disappear as these parameters increase. In addition, the minimum of the effective potential for a rotating embedded BH is higher than that of the corresponding non-rotating case, illustrating a reduced degree of orbital binding.

\subsection{Testing Effective Force Measurement around a rotating and embedded BH model 
}\label{S2-3} 

The motion of a test particle measurement of a BH is controlled by the effective force, which specifies whether the gravitational interaction leads to attraction toward the BH or repulsion away from it. In this context, we analyze the dynamics of particles moving around a rotating and embedded BH, for which the nature of the gravitational interaction depends sensitively on the values of the system parameters. In particular, the combined effects of rotation and embedding modify the radial structure of the gravitational field and, consequently, the force experienced by the particle. Starting from the effective potential defined in Eq. (\ref{Veff-1}), the effective radial force acting on the particle is obtained by taking the derivative of the potential with respect to the radial coordinate, yielding \cite{B1}-\cite{B12}:
\beq
F = - \frac{1}{2}\frac{d V_{eff}}{dr},
\eeq

This definition enables a direct assessment of the direction and magnitude of the gravitational interaction, providing a useful diagnostic for identifying regions of stable and unstable motion.

\begin{figure*}
\centering 
\includegraphics[width=\hsize]{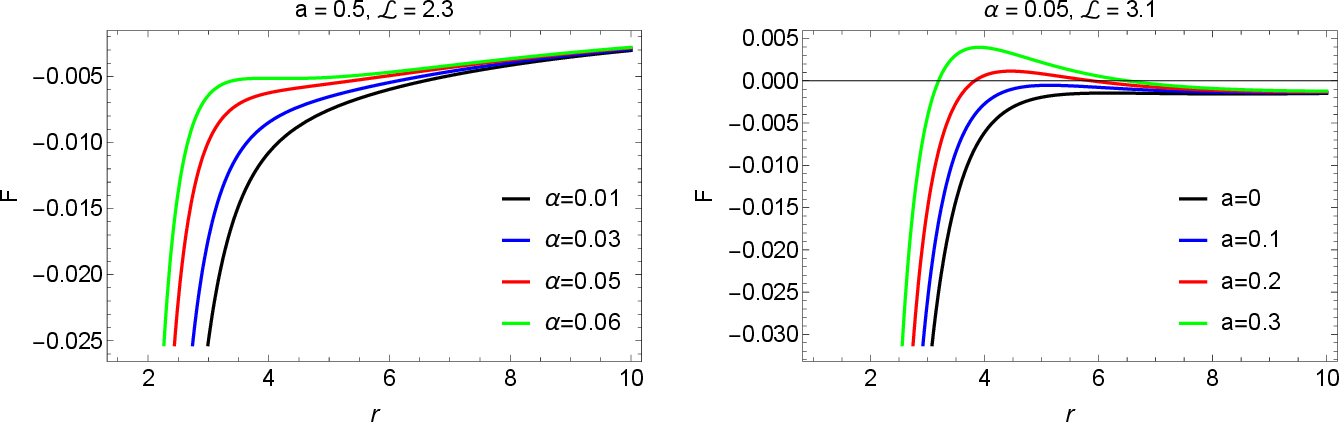}
\caption{Effective force acting on particles orbiting a rotating and embedded BH as a function of the radial coordinate $r$\label{F4}.}\label{figForce}
\end{figure*}

Figure (\ref{figForce}) shows the radial behavior of the effective force for different values of the BH parameters. The left panel focuses on the influence of the embedding parameter $\alpha$, while the right panel illustrates the effect of the rotation parameter $a$. In both cases, a similar qualitative behavior is observed. For relatively small values of $\alpha$ and $a$, the effective force remains weak, indicating a modest gravitational influence on the particle. As either parameter increases, the magnitude of the force is amplified, reflecting the strengthening of the gravitational interaction induced by embedding and rotation effects. For low parameter values, the force is mainly attractive over the considered radial range, and its magnitude increases monotonically as $\alpha$ or $a$ grows. In this case, the enhancement of the force can be attributed to the modification of the spacetime geometry, which alters the balance between centrifugal and gravitational contributions. Also, the results show that the gravitational force generated by a rotating and embedded Kerr BH is stronger than that associated with a standard Kerr BH, illustrating that embedding effects play a non-negligible role in particle dynamics.

\begin{figure*}
\centering 
\includegraphics[width=\hsize]{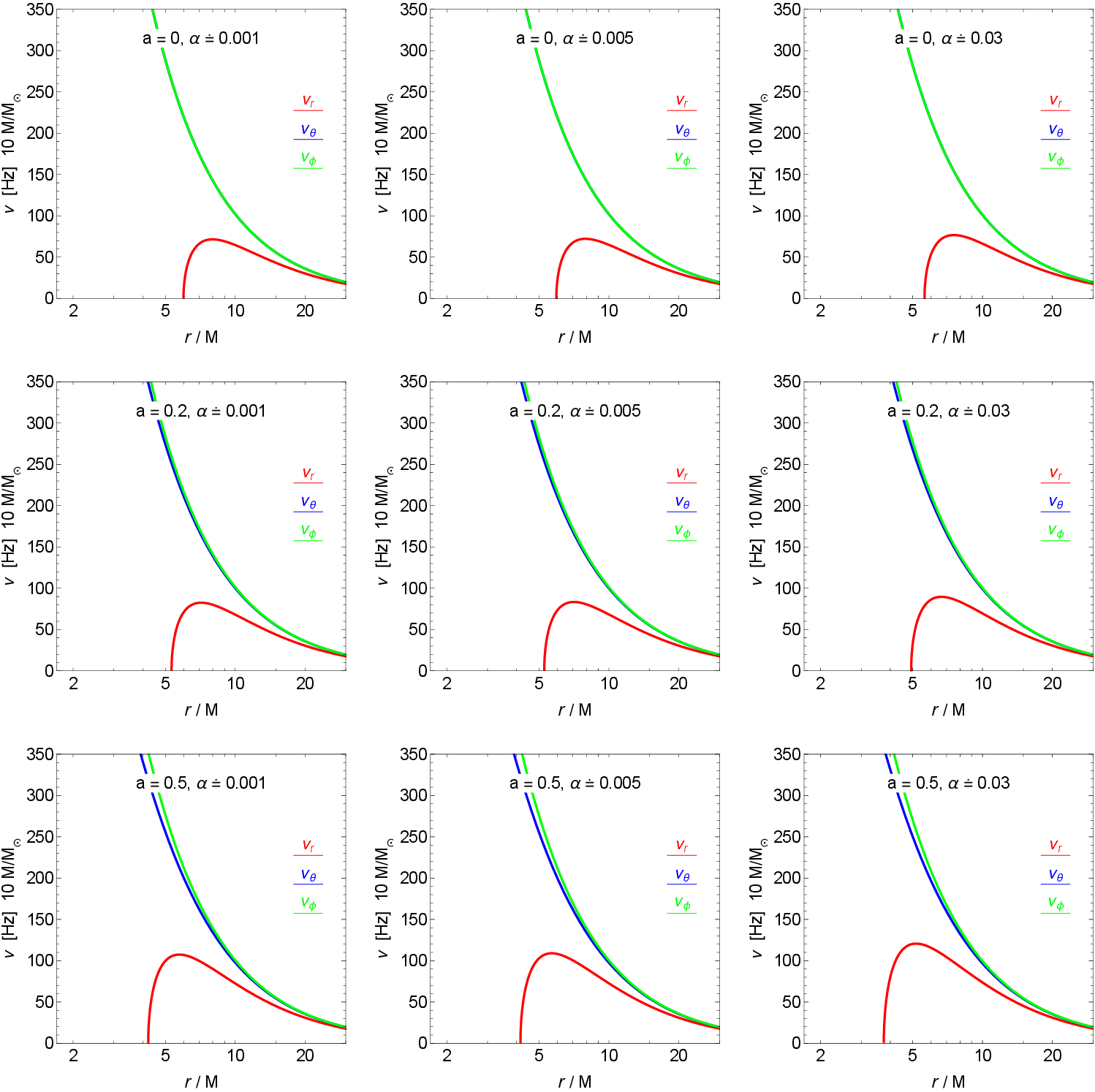}
\caption{Fundamental frequencies of particles moving in the spacetime of a rotating and embedded BH\label{F5}.}\label{figFRQ}
\end{figure*}

\begin{figure*}
\centering 
\includegraphics[width=\hsize]{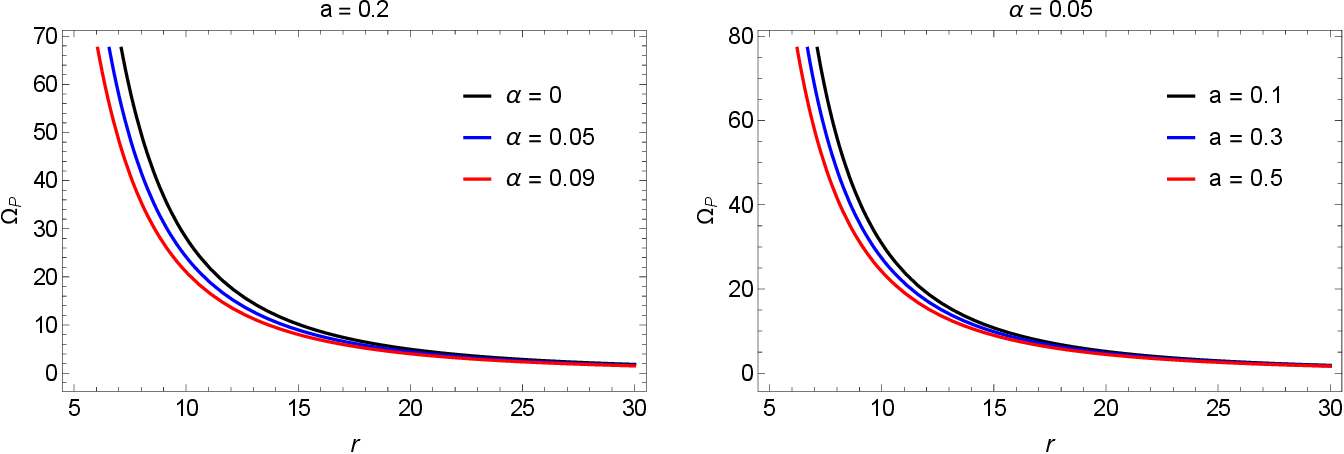}
\caption{Periastron precession frequency for particles orbiting a rotating and embedded BH\label{F6}.}\label{fig_periastron}
\end{figure*}

\begin{figure*}
\centering 
\includegraphics[width=\hsize]{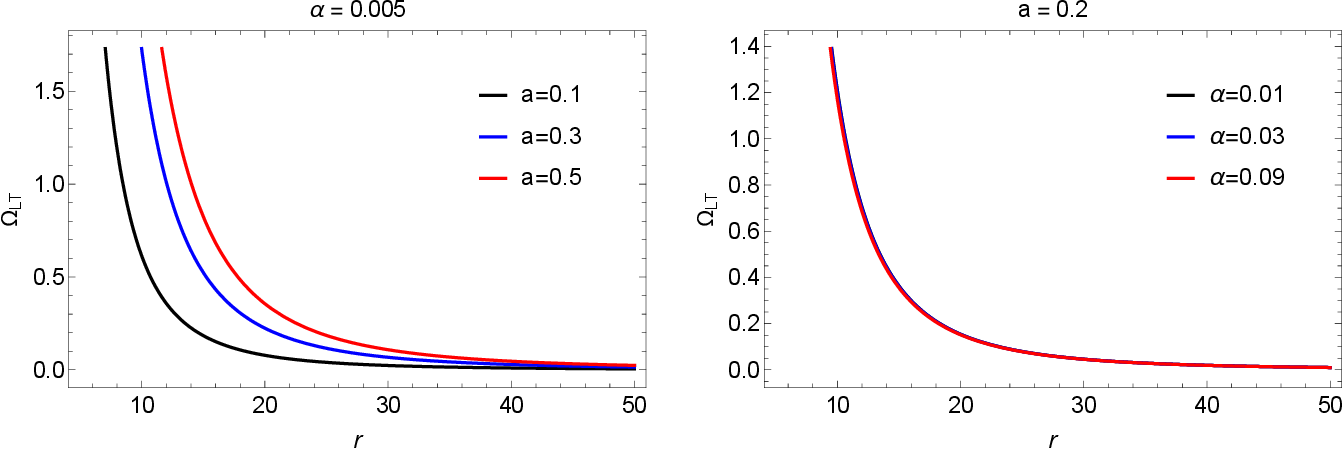}
\caption{Lense-Thirring precession frequency around a rotating and embedded BH\label{F7}.}\label{fig_LT}
\end{figure*}

\section{Testing harmonic oscillation measurement as a perturbation of circular orbits case}\label{S3}

To investigate the oscillatory behavior of neutral test particles, we perform a linear perturbation analysis of the equations of motion around stable circular orbits. Consider a particle initially moving along a stable circular orbit confined to the equatorial plane. If the particle experiences a small displacement from its equilibrium position, the resulting motion can be characterized as epicyclic oscillations, which are effectively linear harmonic in nature.The locally measured epicyclic frequencies are defined as \cite{B1}-\cite{B12}: 
\bea\label{Freq-2}
\omega_{r}^{2} &=&  \frac{-1}{2\, g_{rr}} \frac{\partial^{2} V_{\rm eff} (r, \theta)}{\partial r^{2}},\\\label{Freq-3}
\omega_{\theta}^{2} &=& \frac{-1}{2\, g_{\theta \theta}} \frac{\partial^{2} V_{\rm eff}(r, \theta)}{\partial \theta^{2}},\\\label{Freq-4}
\omega_\phi &=& \frac{\d \phi}{\d \tau}.
\eea
Here, $V_{\rm eff}(r, \theta)$ illustrates the effective potential governing particle motion, and $g_{rr}$ and $g_{\theta\theta}$ are the metric components in the radial and polar directions, respectively. Also, the frequencies $\omega_r$, $\omega_\theta$, and $\omega_\phi$ determine the particle to radial and vertical perturbations, as well as its angular motion. Analysis of these frequencies, including their ratios, provides detailed insight into the geometric structure of epicyclic trajectories and the stability of circular orbits \cite{B1}-\cite{B12}.

In the Newtonian limit, all three frequencies are identical, $\omega_r = \omega_\theta = \omega_\phi$, which implies that particles follow closed elliptical orbits around spherically symmetric masses. In contrast, in the Schwarzschild BH spacetime, the radial frequency is always smaller than the polar and azimuthal frequencies, $\omega_r < \omega_\theta = \omega_\phi$. Also, this discrepancy underlies the relativistic precession of the orbital pericenter. The effect becomes increasingly significant as the orbit approaches the BH, reflecting the influence of strong-field gravitational effects on the particle dynamics. The study of these epicyclic frequencies is therefore essential for understanding relativistic corrections to orbital motion and the behavior of test particles in curved spacetimes.

\subsection{The Exact Frequencies measured by a distant observer}\label{S3-1} 

The locally measured angular frequencies $\omega_{\alpha}$ were derived in Eqs. (\ref{Freq-2})–(\ref{Freq-4}). In order to compare these quantities with measurements performed by a static observer at spatial infinity, it is necessary to introduce the corresponding redshifted angular frequencies $\Omega_{\alpha}$. In this case, the relation between locally defined frequencies and those measured at infinity is governed by the gravitational redshift, which accounts for the time dilation induced by the curved spacetime geometry \cite{B1}-\cite{B12}. Also, the redshifted angular frequencies are therefore defined as
\beq\label{frequencies}
\Omega_{\alpha} = \omega_{\alpha} \frac{\d \tau}{\d t},
\eeq
where $\mathrm{d}\tau/\mathrm{d}t$ illustrate the ratio between the proper time of the particle and the coordinate time measured by a distant observer.

Using the conserved energy $E$ and angular momentum $L$ associated with the stationarity and axial symmetry of the spacetime, given in Eqs. (\ref{EE}) and (\ref{LL}), the gravitational redshift factor can be written explicitly in terms of the metric components as \cite{B1}-\cite{B12}
\begin{equation}
    \frac{\d t}{\d \tau} = - \frac{E g_{\phi \phi} + L g_{t \phi}}{g_{tt} g_{\phi \phi} - g_{t \phi}^2}.
\end{equation}
This expression shows that the observed frequencies depend not only on the orbital constants of motion but also on the frame-dragging effects encoded in the off-diagonal metric component $g_{t\phi}$.

When the oscillation frequencies measured at infinity are expressed in physical units, the corresponding dimensionless quantities must be rescaled by the factor $c^{3}/(GM)$, which naturally arises from restoring physical units in relativistic geometrized expressions. Here, $G$ is the gravitational constant, $c$ is the speed of light, and $M$ denotes the BH mass. As a result, the observable frequencies of neutral particles detected by a distant observer take the form \cite{B1}-\cite{B12}:
\beq\label{nu_rel}
\nu_{j}=\frac{1}{2\pi}\frac{c^{3}}{GM} \, \Omega_{j}[{\rm Hz}].
\eeq
The index $j\in\{r,\theta,\phi\}$ labels the radial, latitudinal, and azimuthal oscillation modes, respectively, as measured at spatial infinity.

For a rotating and embedded BH spacetime, the explicit forms of the redshifted angular frequencies $\Omega_{\alpha}$ are obtained after evaluating the second derivatives of the effective potential around circular orbits. The resulting expressions are given by
\bea\non
\Omega_{r}^{2} &=& \frac{1}{\mathcal{F}} (-2 a^6 \EE^2 (r-12 \alpha )+4 a^5 \EE \LL (r-12 \alpha )-2 a^4 \left(\EE^2 \left(-192 \alpha ^2+3 r^3-6 (6 \alpha +1) r^2+80 \alpha  r\right) \right. \\\non  &+& \left.  \LL^2 (r-12 \alpha )\right)+4 a^3 \EE \LL \left(-192 \alpha ^2+3 r^3-6 (6 \alpha +1) r^2+80 \alpha  r\right) + a^2 (2 \EE^2  (768 \alpha ^3-3 r^5 \\\non &+& 2 (18 \alpha +7) r^4-12 (12 \alpha +1) r^3+144 \alpha  (2 \alpha +1) r^2-576 \alpha ^2 r )-\LL^2 (-384 \alpha ^2+r^4+6 r^3 \\\non &-& 12 (6 \alpha +1) r^2+160 \alpha  r ) )+8 a \EE \LL (-384 \alpha ^3+3 r^5-4 (5 \alpha +2) r^4+(72 \alpha +6) r^3 \\\label{nu_r} &-& 72 \alpha  (2 \alpha +1) r^2+288 \alpha ^2 r )-2 \EE^2 r^4 \left(32 \alpha ^2+r^3-12 \alpha  r^2\right)+3 \LL^2 \left(8 \alpha +r^2-2 r\right)^3)
,\\\non
\Omega_{\theta}^{2} &=& \frac{1}{r^2\mathcal{F}} (\left(a^2+8 \alpha +r^2-2 r\right) (2 a^4 \EE^2 \left(32 \alpha ^2+r^3+(2-4 \alpha ) r^2-16 \alpha  r\right)-8 a^3 \EE \LL (r-4 \alpha )^2 \\\label{nu_theta} &+& a^2 \left(2 \EE^2 r^4 (r-4 \alpha )+\LL^2 \left(64 \alpha ^2+r^4-2 r^3+(8 \alpha +4) r^2-32 \alpha  r\right)\right)+\LL^2 r^4 \left(8 \alpha +r^2-2 r\right) ))
,\\\label{nu_phi}
\Omega_\phi &=& \frac{8 a \alpha -a r+r^2 \sqrt{r-8 \alpha }}{r^4-a^2 (r-8 \alpha )}.
\eea
The function $\mathcal{F}$ ensures the correct normalization of the frequencies and is defined as
\bea
\mathcal{F} &=& r^2 \left(a^2 \EE \left(-8 \alpha +r^2+2 r\right)-2 a \LL (r-4 \alpha )+\EE r^4\right)^2
\eea

Fig. (\ref{figFRQ}) shows the radial behavior of the observable frequencies $\nu_{j}$ associated with small harmonic oscillations of neutral particles orbiting a rotating and embedded BH, as measured by a distant observer, for several values of the spin parameter $a$ and the embedding parameter $\alpha$. The first row corresponds to the non-rotating limit. In the case $a=0$, the radial and vertical frequencies coincide, illustrtaing the spherical symmetry of the spacetime. For nonzero spin, this degeneracy is lifted, leading to a clear separation between the radial and vertical modes due to frame-dragging effects. In addition, increasing the spin parameter shifts the frequency extrema toward smaller radii, closer to the event horizon. A comparable inward shift is observed when $\alpha$ is increased, indicating that the embedding parameter enhances relativistic effects and modifies the effective potential governing particle motion.

\subsection{Exact simulation of Periastron and Lense-Thirring precession Measurement under rotating and embedded BH}\label{S3-2} 

In this subsection, we examine the Lense-Thirring precession and the periapsis precession of a neutral test particle moving around a rotating and embedded BH, with its trajectory slightly displaced from the equatorial plane ($\theta=\pi/2$). In this context, the particle is assumed to experience small perturbations around an otherwise circular and stable orbit. As a consequence, the motion can be decomposed into oscillations about an equilibrium radius, which are described by a characteristic radial frequency $\Omega_r$. Also, this radial oscillatory behavior plays a central role in determining the periapsis precession, since it quantifies the deviation from a closed Keplerian orbit induced by relativistic effects.

The Lense-Thirring precession frequency $\Omega_{LT}$ is defined as the difference between the azimuthal (orbital) frequency $\Omega_\phi$ and the latitudinal frequency $\Omega_\theta$, reflecting the dragging of inertial frames caused by the BH rotation. Also, the periapsis (periastron) precession frequency $\Omega_P$ is given by the difference between the azimuthal frequency $\Omega_\phi$ and the radial frequency $\Omega_r$, which measures the advance of the orbital periapsis per revolution due to spacetime curvature and rotation effects \cite{B1}-\cite{B12}. These definitions are written as
\beq
\Omega_{P} = \Omega_{\phi} - \Omega_{r},
\eeq
and
\beq
\Omega_{LT} = \Omega_{\phi} - \Omega_{\theta}.
\eeq
Fig. (\ref{fig_periastron}) presents the behavior of the periapsis precession frequency for a rotating and embedded BH for several choices of the BH parameters. In this case, both parameters affect $\Omega_P$ in a similar qualitative manner, illustrtaing that they contribute comparably to the relativistic corrections governing the orbital dynamics. More specifically, an increase in the rotation parameter $a$ results in a systematic decrease of the periapsis precession frequency. In addition, a similar reduction of $\Omega_P$ is observed when the black hole parameter $\alpha$ is increased, showing that this parameter also weakens the precessional effect associated with radial oscillations.

The dependence of the Lense-Thirring precession frequency on the BH parameters is shown in Fig. (\ref{fig_LT}). The first column illustrates the variation of $\Omega_{LT}$ for different values of the rotation parameter $a$, whereas the second column displays the corresponding behavior for different values of the parameter $\alpha$. In this case, the results clearly indicate that the BH rotation enhances the Lense-Thirring precession, consistently with the interpretation of frame-dragging effects. By contrast, increasing the parameter $\alpha$ leads to a reduction of $\Omega_{LT}$, implying that this parameter counteracts the rotational contribution to the precession of the orbital plane.

\section{General Relativistic Hydrodynamic Simulations of Bondi-Hoyle-Lyttleton Accretion onto Rotating Embedded Black Holes}\label{S4} 

The BHL mechanism reveals the physical consequences of the interaction between matter and a black hole within a strong gravitational field and contributes to the measurement of astrophysical phenomena that can be observed observationally \cite{Koyuncu:2014nga,Donmez:2022dze,Donmez:2026yvm}. This is because the BHL mechanism naturally generates a shock-cone structure around the black hole, enabling the capture and confinement of physical processes occurring within this cone. Unlike the thin-disk approach, which assumes nearly Keplerian motion and exhibits axisymmetric behavior, the BHL accretion mechanism produces observable radiation by capturing supersonic matter through gravitational attraction. This observable radiation carries information about the interaction between matter and the black hole, while simultaneously providing insights into the geometric structure of the black hole spacetime and the physical properties of the black hole itself. Moreover, the shock-cone structure formed in the strong gravitational-field regime enables not only the interpretation of QPO frequencies observed in X-ray binary (XRB) and active galactic nucleus (AGN) sources but also offers clues about black hole shadows observed by the Event Horizon Telescope (EHT). For these reasons, even when restricted to the equatorial plane, it is important to analyze the morphology and dynamics of accretion resulting from the accumulation of matter onto the black hole via the BHL mechanism, as well as the response of this dynamical structure to variations in spacetime parameters.

Based on these considerations, we perform systematic numerical modeling in order to both contribute to the understanding of the physical properties of the black hole, such as its spin and mass, and to propose physical mechanisms for the observed QPOs. In numerical models, the general relativistic hydrodynamic (GRD) equations are solved using the black hole spin and spacetime parameters listed in Table \ref {ST_param}, and the effects of spacetime geometry and black hole rotation on the shock cone formed around rapidly and moderately rotating black holes are investigated. In this framework, the influence of the spacetime parameters and their interaction with the spin parameter on the dynamics of the shock cone are examined in a systematic manner. When selecting the spacetime parameters in Table \ref {ST_param}, only configurations admitting horizons and therefore not producing naked singularities are considered. For each model listed in Table \ref {ST_param}, numerical simulations are performed to reveal in detail both the frame-dragging effects arising from black hole rotation and the embedding corrections affecting the spacetime geometry, as well as their combined influence on the shock-cone dynamics. In addition, the numerical results obtained are discussed in detail in terms of their consistency with theoretical predictions based on test-particle motion and with observational findings.

\begin{table}[h]
\centering
\caption{Horizons and qualitative deviation from the Kerr spacetime for rotating and embedded black holes with $M=1$. The table combines the horizon-admitting case at $a=0.9$ and $\alpha=0.01$ with a sequence of increasing deformation parameter $\alpha$ for $a=0.5$. Increasing $\alpha$ leads to progressively stronger departures from Kerr geometry, reflected in modified horizon locations and increasingly unstable BHL accretion dynamics.}
\label{ST_param}
\begin{tabular}{c c c c c}
\hline
\hline
\hspace{0.5cm}\textbf{Model}\hspace{0.5cm} & \hspace{0.5cm}$a$\hspace{0.5cm} & \hspace{0.5cm}$\alpha (M^2)$ \hspace{0.5cm}& \hspace{0.5cm}Horizons $(r/M)$\hspace{0.5cm} & \hspace{0.5cm} Deviation from Kerr \\
\hline\hline
Kerr-09 & 0.9 & 0.0 & $r_{-}=0.5641,\;\; r_{+}=1.4358$ &  Kerr \\
A1 & 0.9 & 0.01 & $r_{-}=0.6683,\;\; r_{+}=1.3317$ & Very close to Kerr \\
Kerr-05 & 0.5 & 0.0 & $r_{-}=0.1339,\;\; r_{+}=1.8660$ &  Kerr \\
B1 & 0.5 & 0.01 & $r_{-}=0.1815,\;\; r_{+}=1.8185$ & Small deviation  \\
B2 & 0.5 & 0.03 & $r_{-}=0.2859,\;\; r_{+}=1.7141$ & Moderate deviation \\
B3 & 0.5 & 0.05 & $r_{-}=0.4084,\;\; r_{+}=1.5916$ & Strong deviation  \\
B4 & 0.5 & 0.07 & $r_{-}=0.5641,\;\; r_{+}=1.4359$ & Very strong deviation \\
B5 & 0.5 & 0.09 & $r_{-}=0.8268,\;\; r_{+}=1.1732$ & Extreme deviation  \\
\hline
\hline
\end{tabular}
\end{table}

\subsection{Morphological Transitions of the Shock Cone and Angular Density Modulation}\label{S4_1} 

\begin{figure*}
  \center
   \psfig{file=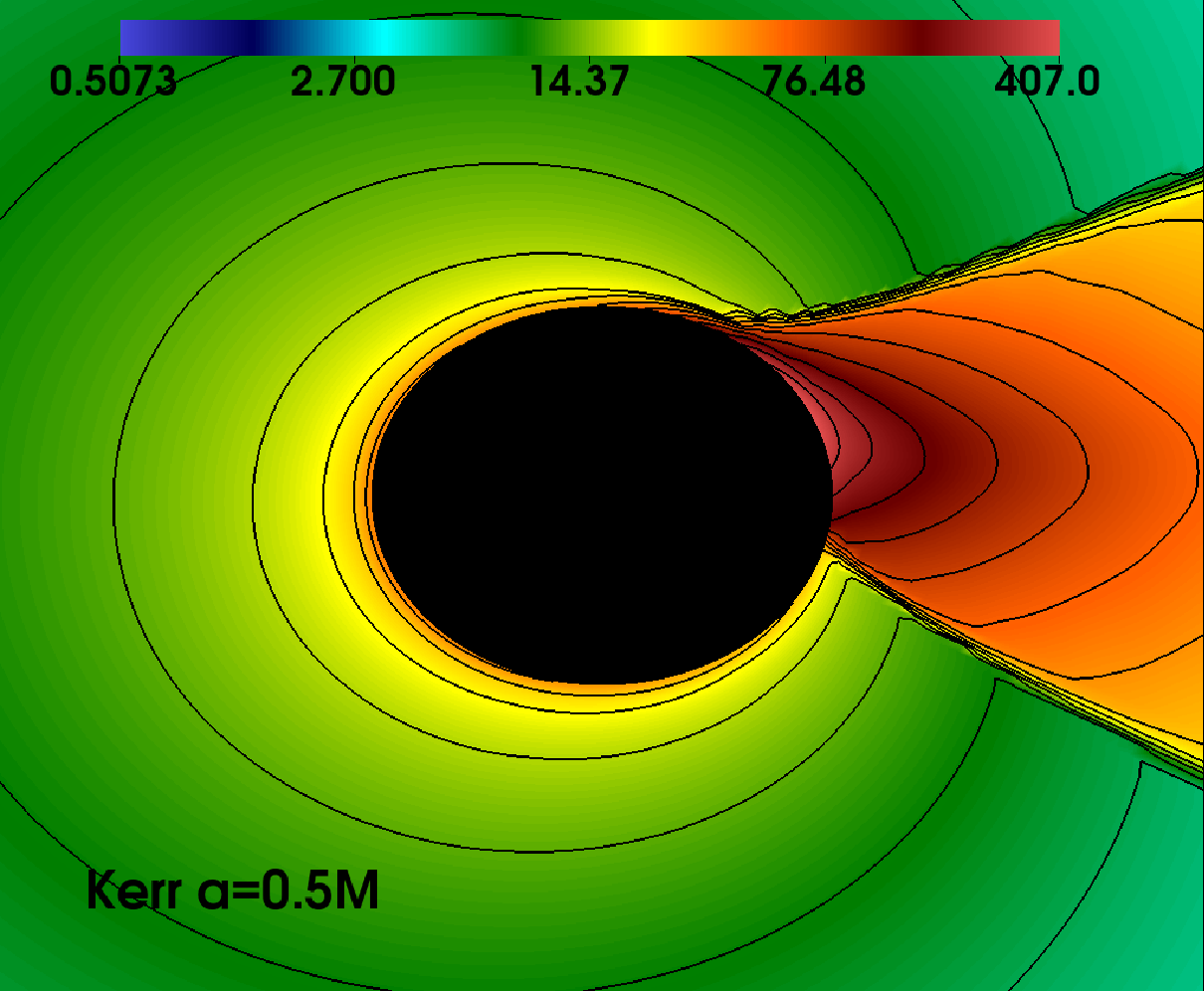,width=7.5cm,height=7.0cm} 
   \psfig{file=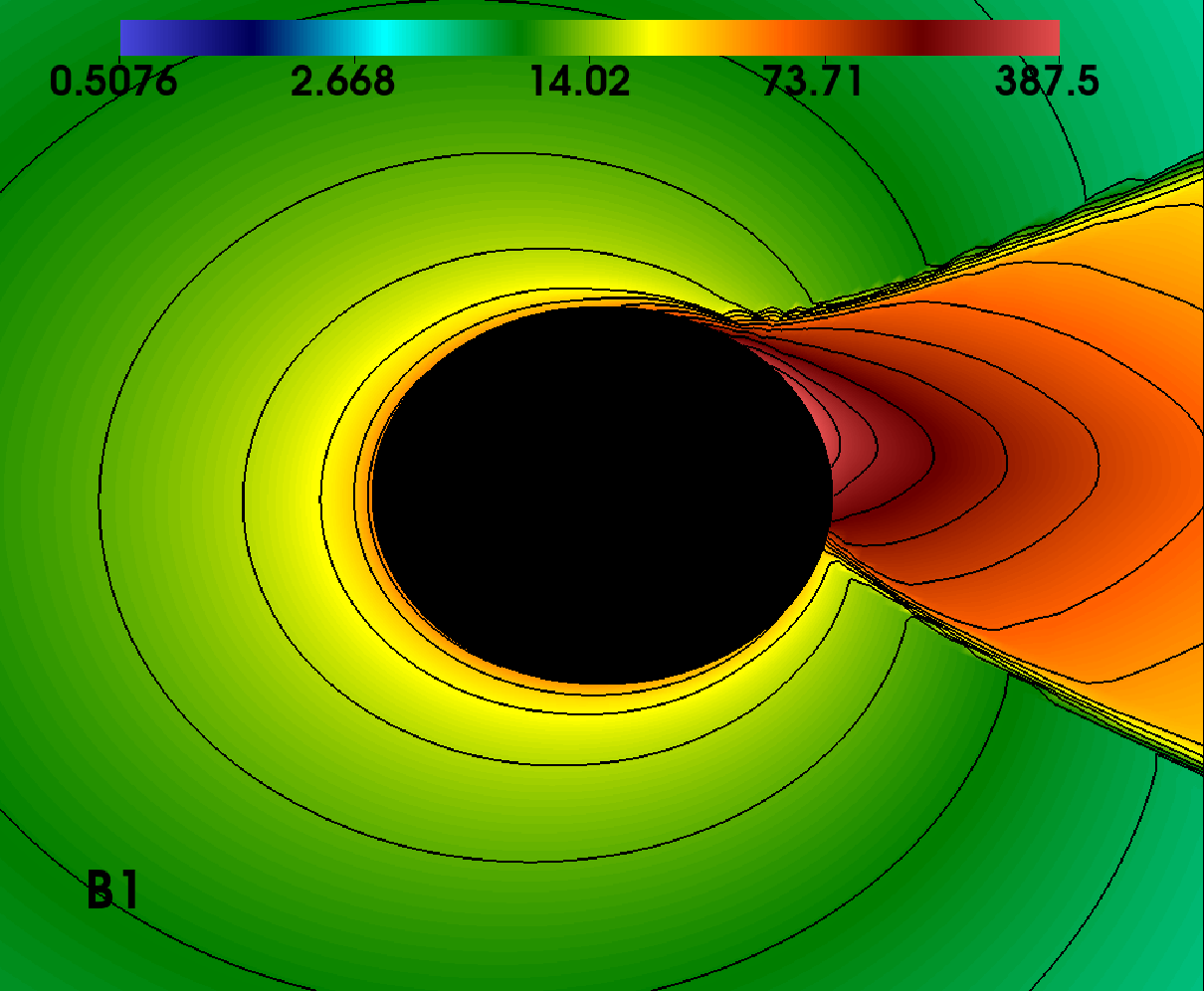,width=7.5cm,height=7.0cm} \\
   \psfig{file=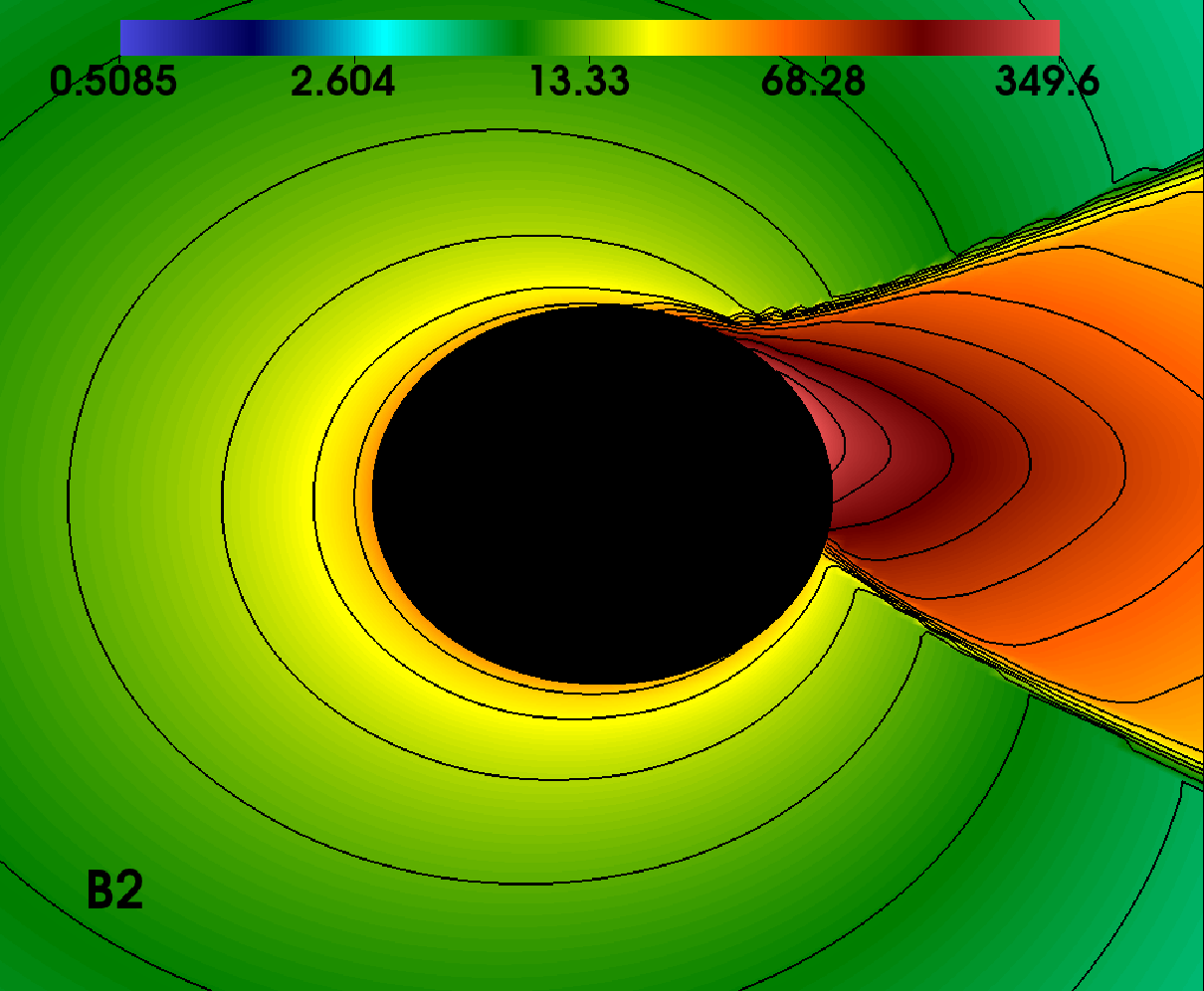,width=7.5cm,height=7.0cm} 
   \psfig{file=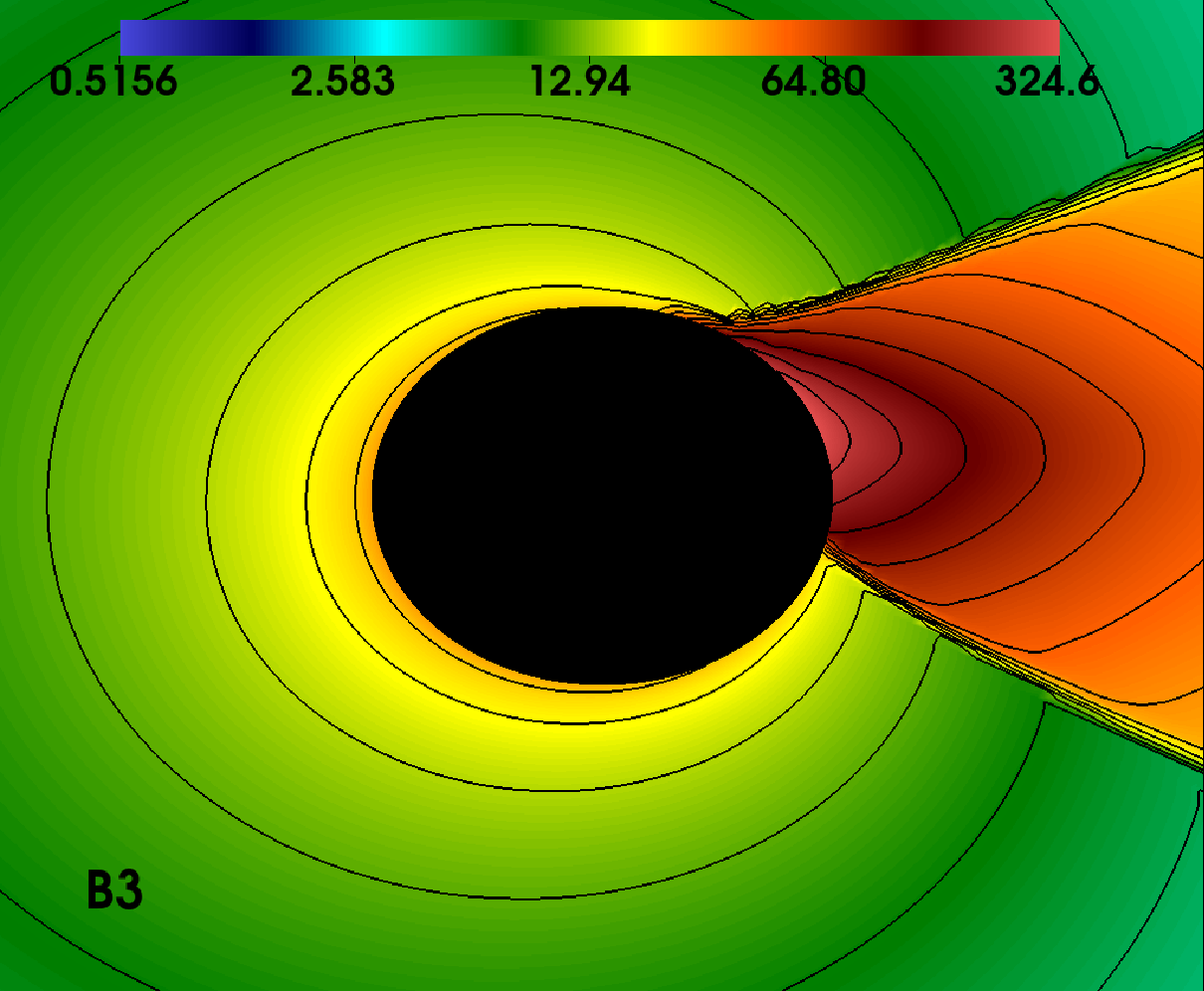,width=7.5cm,height=7.0cm} \\ 
   \psfig{file=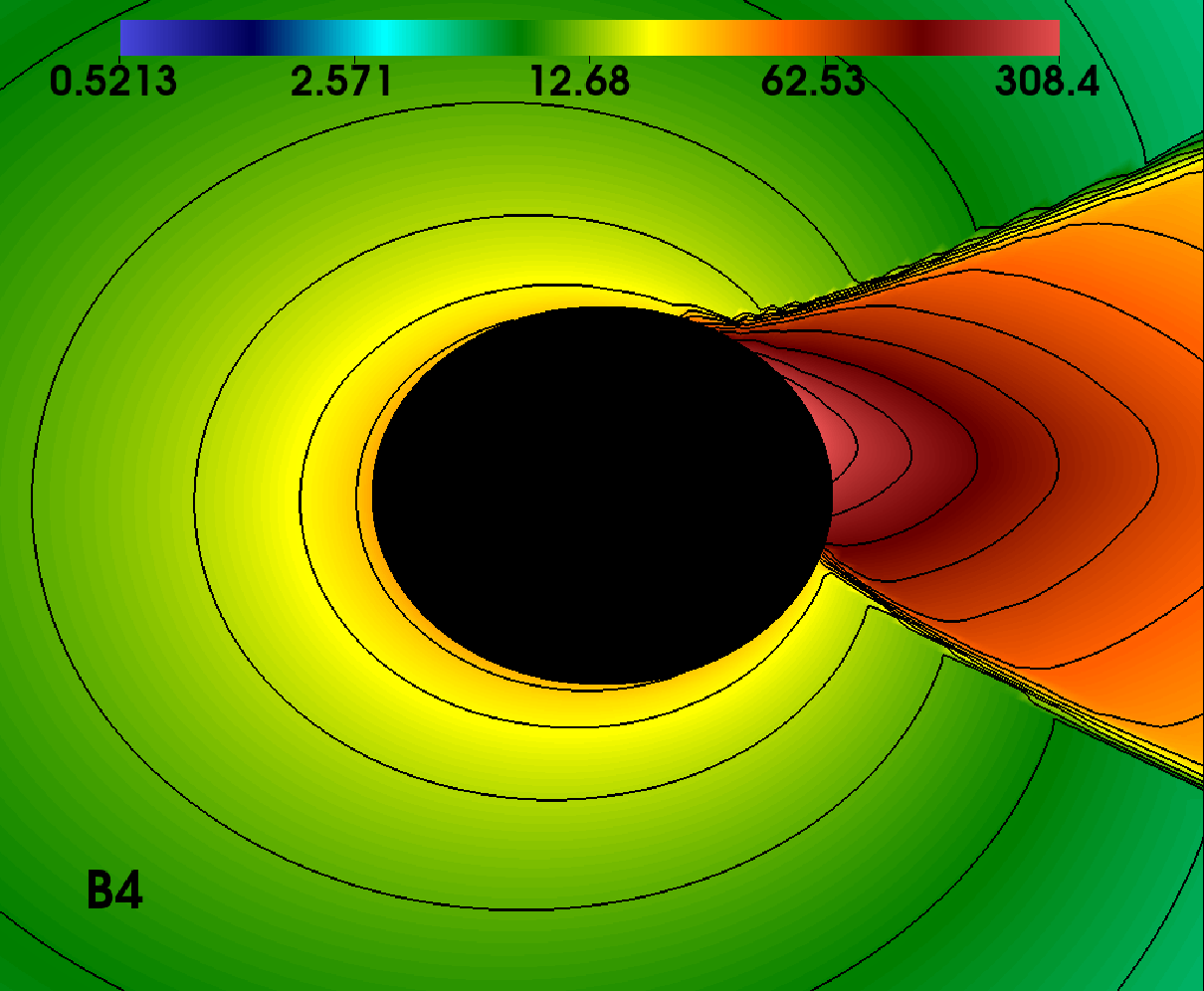,width=7.5cm,height=7.0cm} 
   \psfig{file=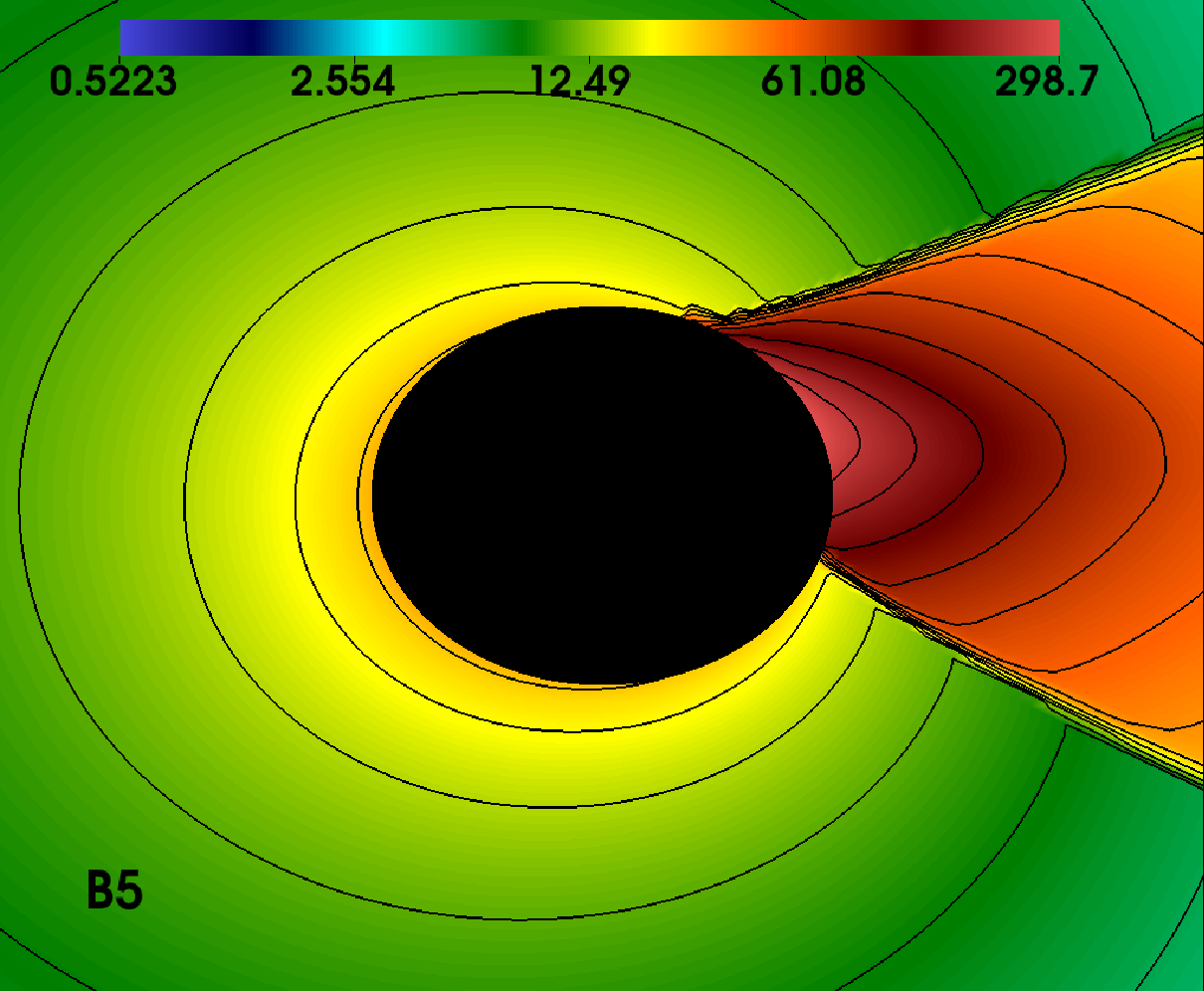,width=7.5cm,height=7.0cm}    
\caption{ The rest-mass density distribution in the equatorial plane is shown using color and contour plots, illustrating the structure of the shock cone formed via the BHL accretion mechanism around a Kerr black hole with rotation parameter $a=0.5$ and around rotating embedded black holes corresponding to the B models listed in Table \ref{ST_param}.  The top-left panel shows the Kerr case, while panels B1-B5 display the deformations arising from increasing values of the embedding parameter $\alpha$, where the deformation of the shock cone, variations in its opening angle, and the suppression of post-shock compression indicate increasing deviations from the Kerr geometry as $\alpha$ increases. 
}
\label{color_plot}
\end{figure*}

The accretion of matter around embedded black holes is strongly influenced by the strong gravitational-field regime. This is because the embedding parameter $\alpha$ directly affects the physical parameters in the region close to the black hole horizon, and therefore the resulting changes are felt in a pronounced manner. In order to visualize this effect, Fig.\ref {color_plot} presents the structure of the shock cone formed via the  BHL mechanism in the equatorial plane around a moderately rotating black hole, as defined by the B models listed in Table \ref {ST_param}. The rest-mass density distribution is shown using color and contour plots, with a zoom applied to the region close to the black hole horizon.  In Fig.\ref {color_plot}, the upper-left panel in the first row corresponds to the Kerr case ($\alpha = 0$), while the remaining panels illustrate the effects in the strong gravitational-field region for increasing model indices associated with increasing values of the embedding parameter $\alpha$. In this way, the influence of the embedding parameter on both the frame-dragging effect generated by the black hole rotation and the morphology of the resulting shock cone is clearly revealed.

In the upper-left panel of   Fig.\ref {color_plot} , corresponding to the Kerr case, a strongly collimated shock structure is observed, together with a pronounced compression in the post-shock region. The density contours are tightly packed downstream, indicating efficient gravitational focusing and a clear imprint of frame-dragging effects associated with the moderate spin of the black hole. This configuration characterizes the strong-field behavior of the Kerr spacetime and is therefore used as a reference for understanding the effects observed in the B models.

\begin{figure*}
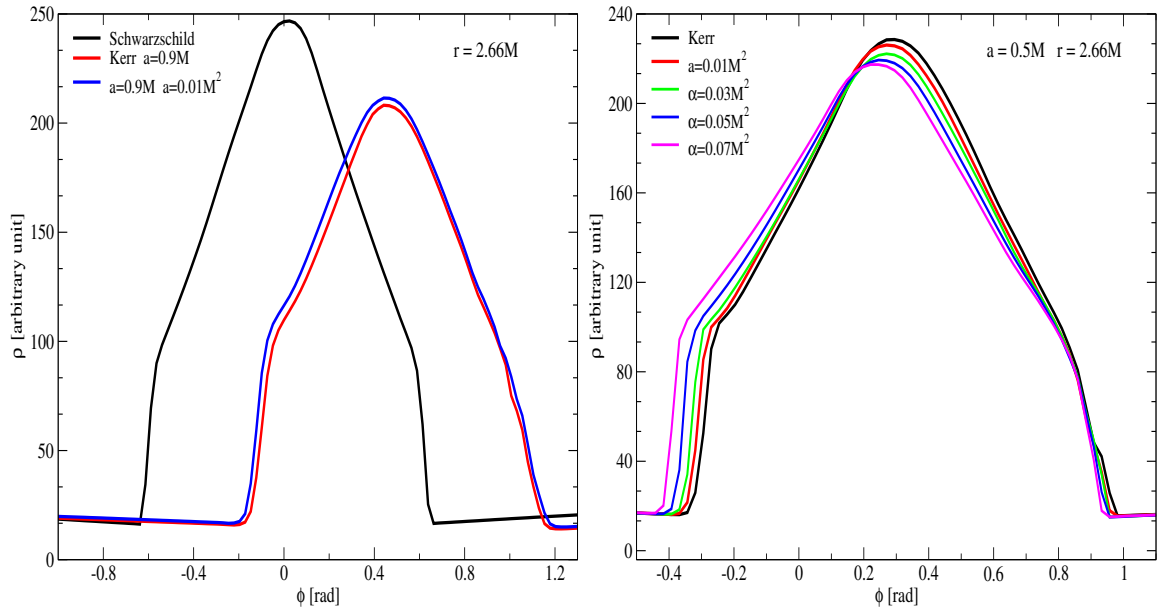

  \center
   \psfig{file=den_phi_r266_FD.eps,width=7.5cm,height=8.0cm} 
   \psfig{file=den_phi_a05_r266.eps,width=7.5cm,height=8.0cm} \\   
\caption{ The azimuthal variation of the rest-mass density of the shock cone and the ambient medium at $r=2.66M$ is presented, where the left panel compares the Schwarzschild and Kerr black hole models, while the right panel shows the changes induced by increasing values of the embedding parameter $\alpha$ for $a=0.5M$, and the distortion of the shock cone together with the asymmetric structure of the density peak demonstrates the effects of rotation and embedding parameters on the shock-cone morphology.
}
\label{Den_Azim_FD}
\end{figure*}

For the B1 model listed in Table \ref {ST_param}, where the embedding parameter is small ($\alpha = 0.01M^{2}$), the physical structure of the shock cone in the strong gravitational-field region remains very similar to that of the Kerr model. As seen in the right panel of Fig.\ref {Den_Azim_FD}, the opening angle of the shock cone increases only by a very small amount. Although the frame-dragging effect is theoretically expected to weaken slightly, this reduction is not easily distinguishable when compared with the Kerr case. With increasing values of the embedding parameter in models B2 and B3, a moderate deviation from the Kerr configuration is observed when compared with the Kerr model. This deviation is again visible in the right panel of Fig.\ref {Den_Azim_FD},. By comparing the color legends of the density distributions in Fig.\ref {color_plot}, it is evident that the density gradient becomes smoother, indicating a weakening of the effective gravitational focusing. Even though the same spin value as in the Kerr and B1 cases is used, the influence of frame dragging is observed to begin to decrease. For larger embedding parameters, corresponding to models B4 and B5, a clear deviation from the Kerr case is observed. The density in the post-shock region is more strongly suppressed and the compression is significantly reduced. In the strong gravitational-field region, the weakening of the frame-dragging effect, which is responsible for the Lense--Thirring epicyclic frequencies, becomes apparent. This leads to variations in the associated QPO frequencies. This indicates that the embedding parameter counteracts the rotational effects by modifying the effective gravitational potential and redistributing matter within the shock cone.

Fig.\ref {Den_Azim_FD} presents the one-dimensional azimuthal variation of the rest-mass density at $r = 2.66M$ in the strong gravitational-field region for all models listed in Table \ref {ST_param}. The dynamic structure of the shock cone, its maximum density, and the density of the ambient medium are clearly visible. In the left panel of Fig.\ref {Den_Azim_FD}, the dynamical structure of the shock cone formed around a Schwarzschild black hole and around Kerr and embedded black holes with a rotation parameter of $a = 0.9M$ is shown. In the Schwarzschild case, the presence of a symmetric and sharply bounded density peak around $\phi = 0$ indicates the absence of frame-dragging effects.  In contrast, for the Kerr model with the rotation parameter $a = 0.9M$, a clear asymmetry is observed, and the location of the maximum density is significantly shifted to approximately $\phi = 0.6\,\mathrm{rad}$. This behavior is a direct consequence of strong frame-dragging effects. For an embedded black hole with the same rotation parameter and a small embedding parameter value of $\alpha = 0.01M^{2}$, a deviation from the Kerr case is observed, with the density profile shifting slightly toward the Schwarzschild configuration. This clearly indicates a weakening of the frame-dragging effect. This demonstrates that even small embedding corrections can significantly alter the strong-field accretion structure by reducing the rotational imprint on the flow.

In the right panel of Fig.\ref {Den_Azim_FD}, as discussed previously, the azimuthal variation of the rest-mass density is shown for Kerr and embedded black hole models with a moderate rotation parameter of $a = 0.5M$. As clearly seen, in the Kerr case the maximum density occurs around $\phi \simeq 0.3\,\mathrm{rad}$, whereas with increasing values of the embedding parameter $\alpha$, this peak gradually shifts toward $\phi = 0$. For model B4, which corresponds to $\alpha = 0.07$, the maximum density is located at approximately $\phi \simeq 0.22\,\mathrm{rad}$. This clearly confirms that increasing the values of the embedding parameter significantly reduces the strength of the frame-dragging effect. The weakening of frame dragging together with the widening of the shock cone is expected to lead to variations in the QPO frequencies generated by oscillation modes captured by such physical mechanisms. In conclusion, Fig.\ref {Den_Azim_FD}  clearly demonstrates that embedding effects can produce observable imprints on QPOs generated by accretion processes occurring in the strong gravitational-field region close to the black hole.

\subsection{Time-Dependent Accretion Dynamics and Variability Signatures }\label{S4_2} 

Understanding the behavior of the mass accretion rate around black holes provides important information for interpreting the luminosity produced in different regions surrounding the black hole and for understanding the frequencies of possible QPO oscillations. The mass accretion rate offers detailed insight into the strength of the interaction between the black hole and the accreting matter, as well as into the instability, turbulence, or stability of the processes that develop in the flow. Therefore, investigating the mass accretion rate at different radial locations around the black hole represents one of the most important physical variables for comparing numerical models with observations. For this reason, in this section we compute the mass accretion rates for different black hole spin parameters and embedding parameters and analyze the physical behavior that emerges at different radial positions.

\begin{figure*}
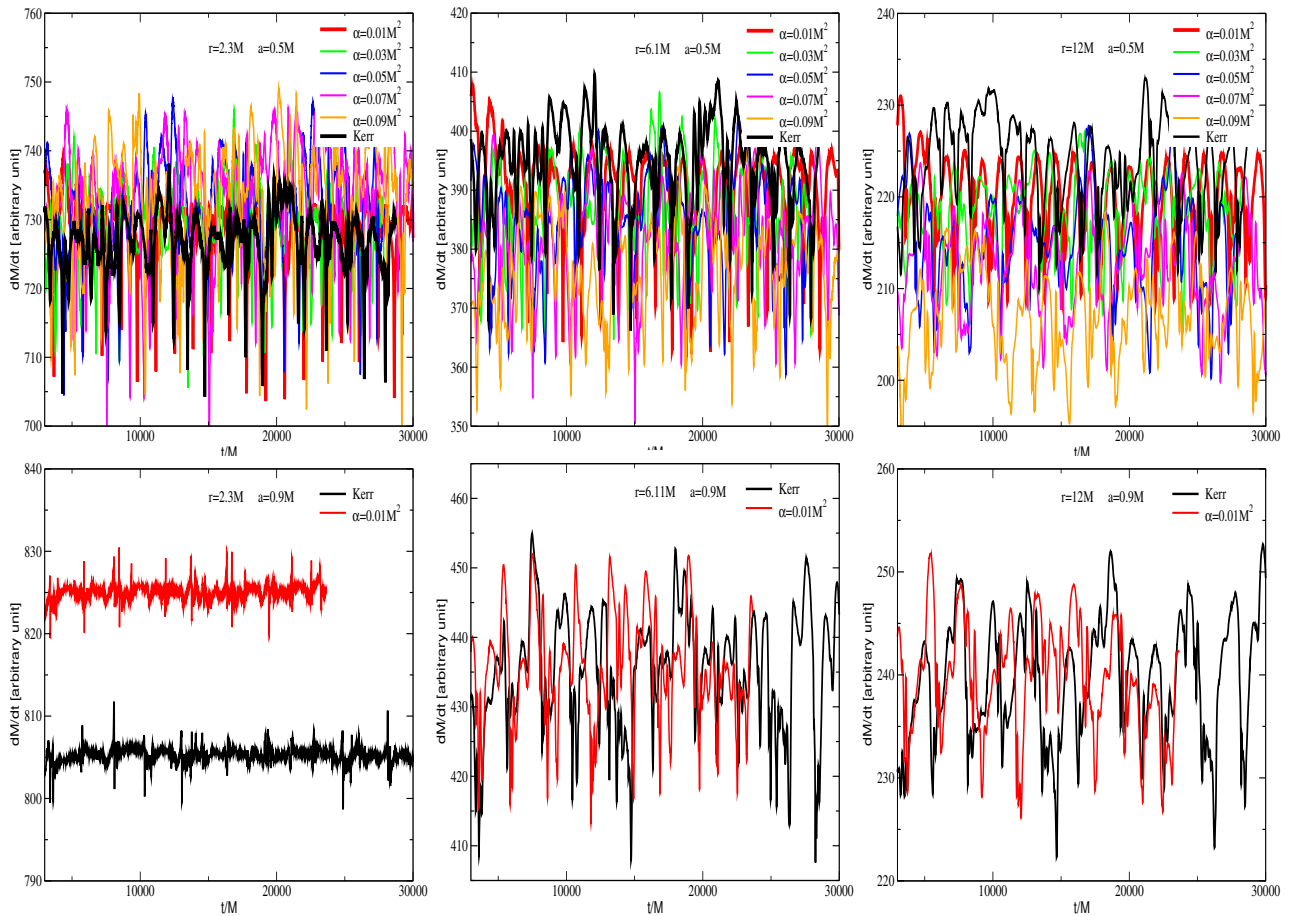

  \center
   \psfig{file=acc_rate_a05_r23.eps,width=5.5cm,height=6.0cm} 
   \psfig{file=acc_rate_a05_r61.eps,width=5.5cm,height=6.0cm}    
   \psfig{file=acc_rate_a05_r12.eps,width=5.5cm,height=6.0cm} \\
   \psfig{file=acc_rate_a09_r23.eps,width=5.5cm,height=6.0cm} 
   \psfig{file=acc_rate_a09_r61.eps,width=5.5cm,height=6.0cm}    
   \psfig{file=acc_rate_a09_r12.eps,width=5.5cm,height=6.0cm}   
\caption{The time evolution of the mass accretion rate is plotted at different radial locations and for different black hole parameters, where the upper panels show the $\alpha$-dependent variations and oscillatory behavior of the accretion rate for $a=0.5M$ at $r=2.3M$, $r=6.11M$, and $r=12M$ from left to right, while the lower panels present the corresponding cases for the same radii with rotation parameter $a=0.9M$. For both rotation parameters, the embedded black hole models are shown in comparison with the Kerr models, and the enhanced fluctuations together with long-term strong variations confirm that the dynamical variability of the accretion flow increases relative to the Kerr spacetime.  
}
\label{mass_acc_a1}
\end{figure*}

\begin{figure*}
  \center
   \psfig{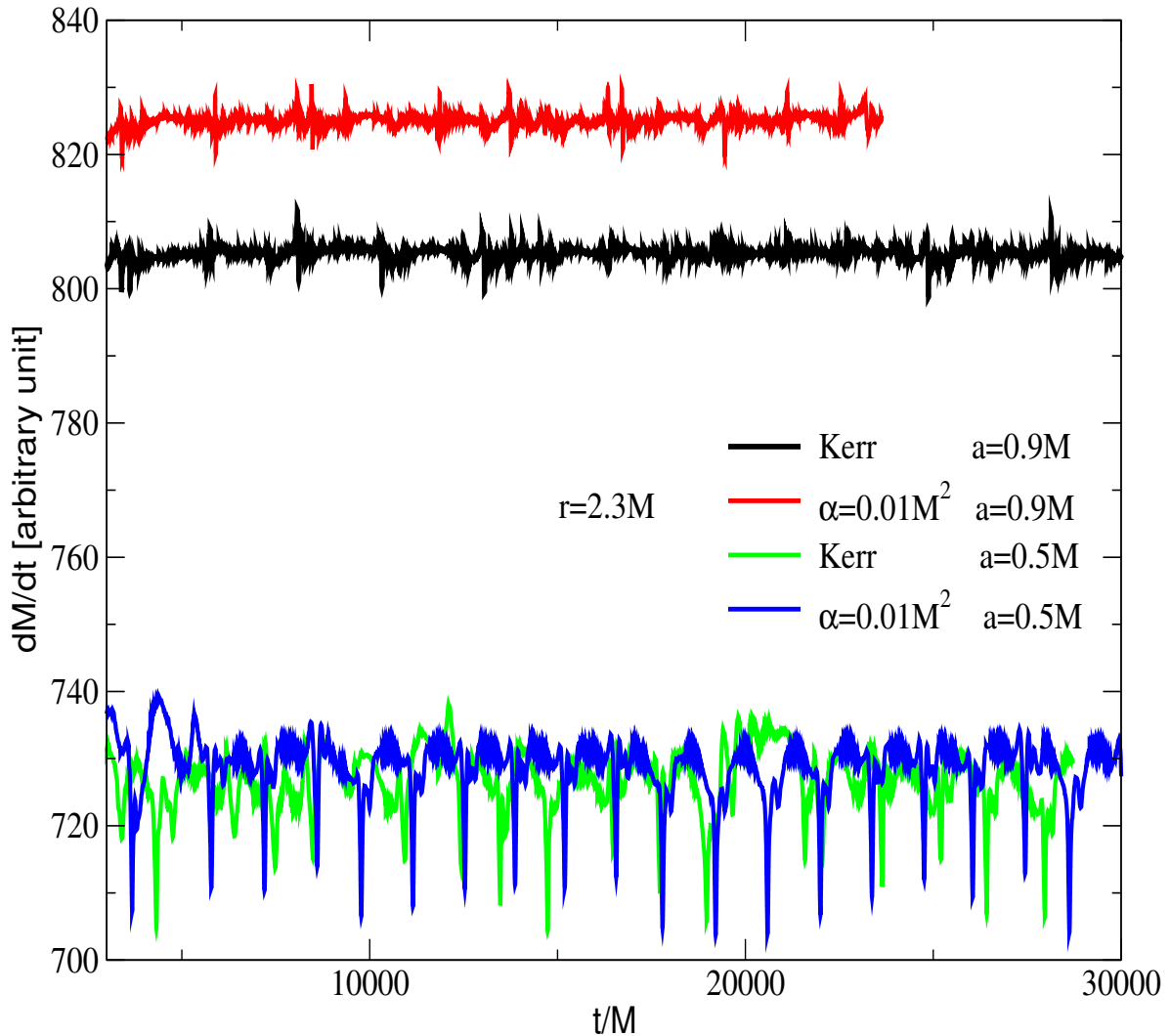}
\caption{  A direct comparison of the mass accretion rate at $r=2.3M$ around Kerr and embedded black holes is presented for different spin values $a=0.5M$ and $a=0.9M$, where even in the case of a small perturbation with $\alpha=0.01M^{2}$, the deviation from the Kerr spacetime in the strong gravitational-field region is clearly distinguishable, indicating that the parameter $\alpha$ significantly modifies the gravitational potential in the vicinity of the black-hole horizon. 
}
\label{mass_acc_A1B1}
\end{figure*}

\begin{figure*}
  \center
  \psfig{file=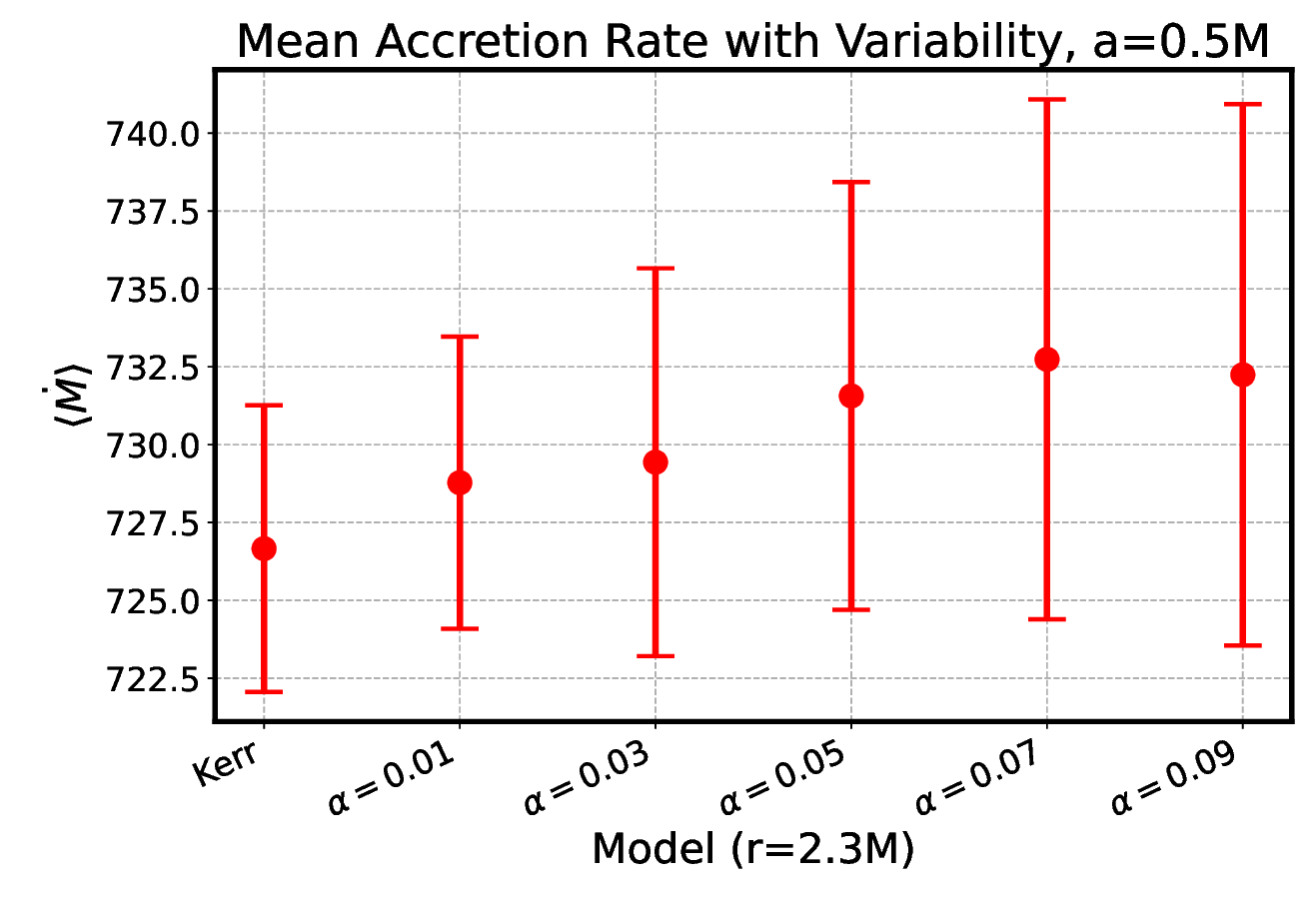,width=5.5cm,height=6.0cm}    
   \psfig{file=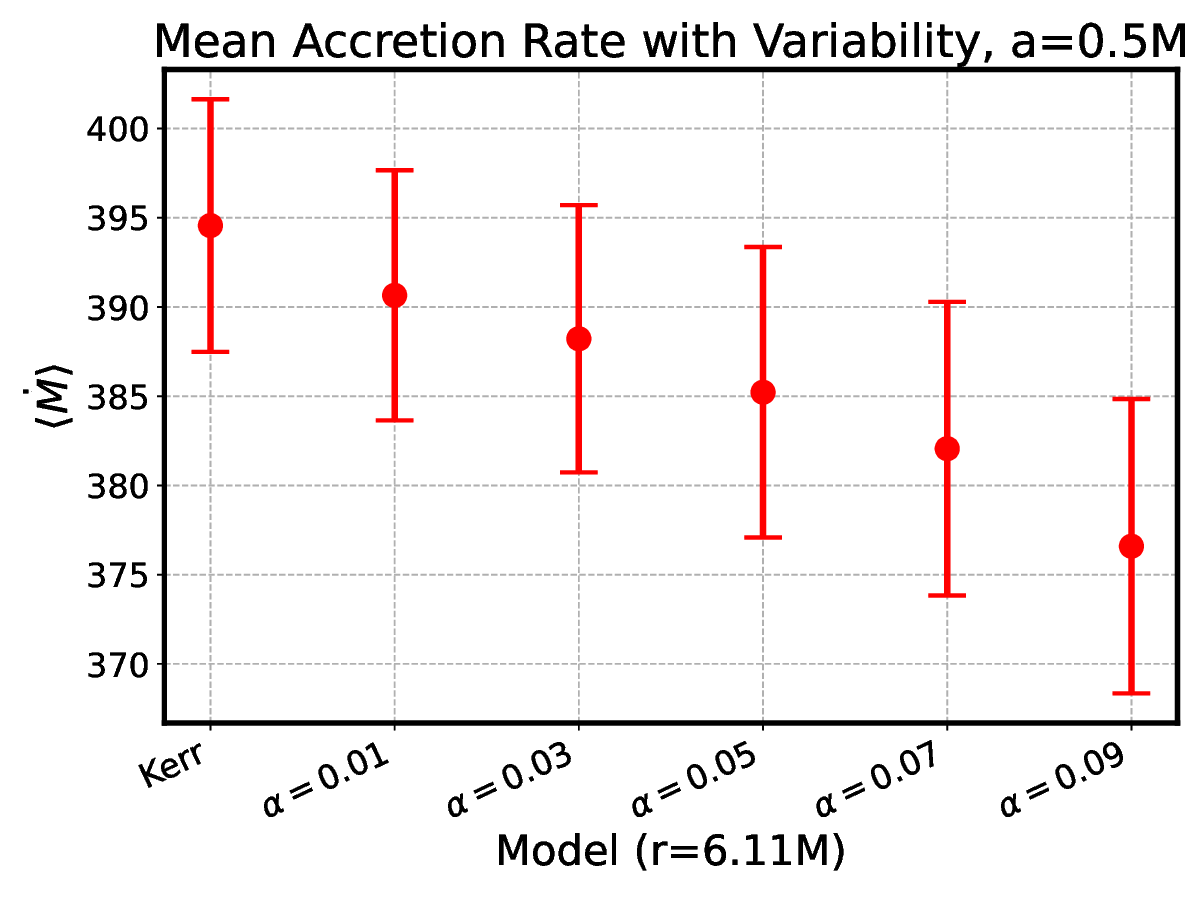,width=5.5cm,height=6.0cm}  
   \psfig{file=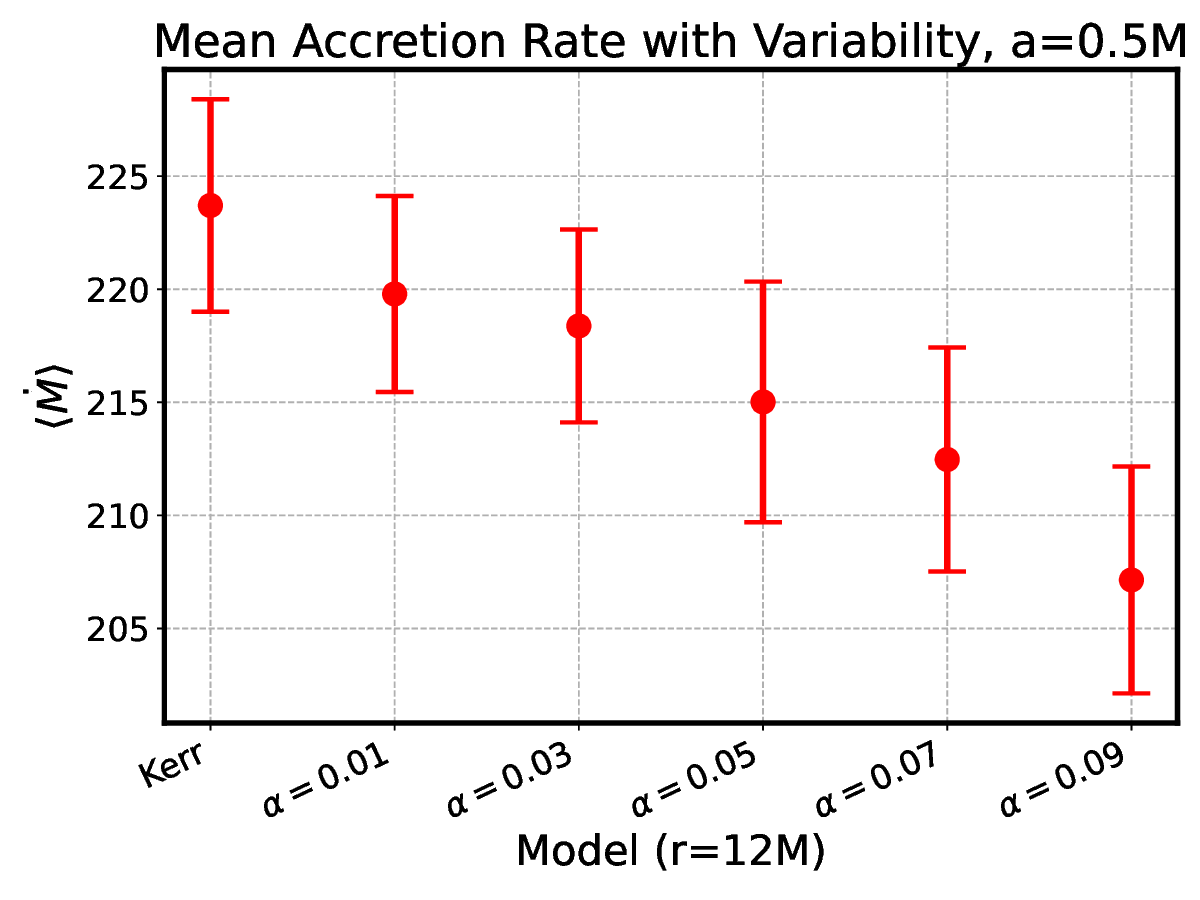,width=5.5cm,height=6.0cm}\\  
   \psfig{file=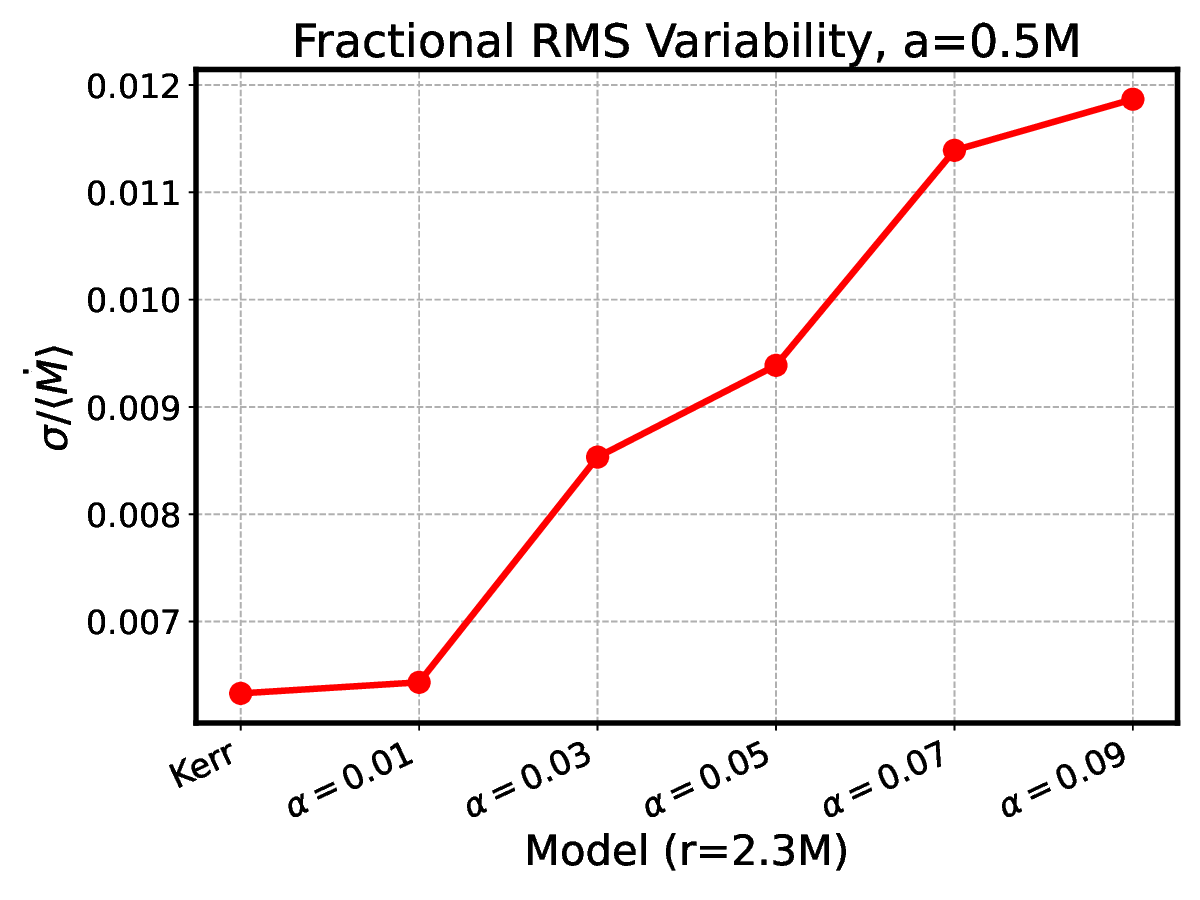,width=5.5cm,height=6.0cm} 
   \psfig{file=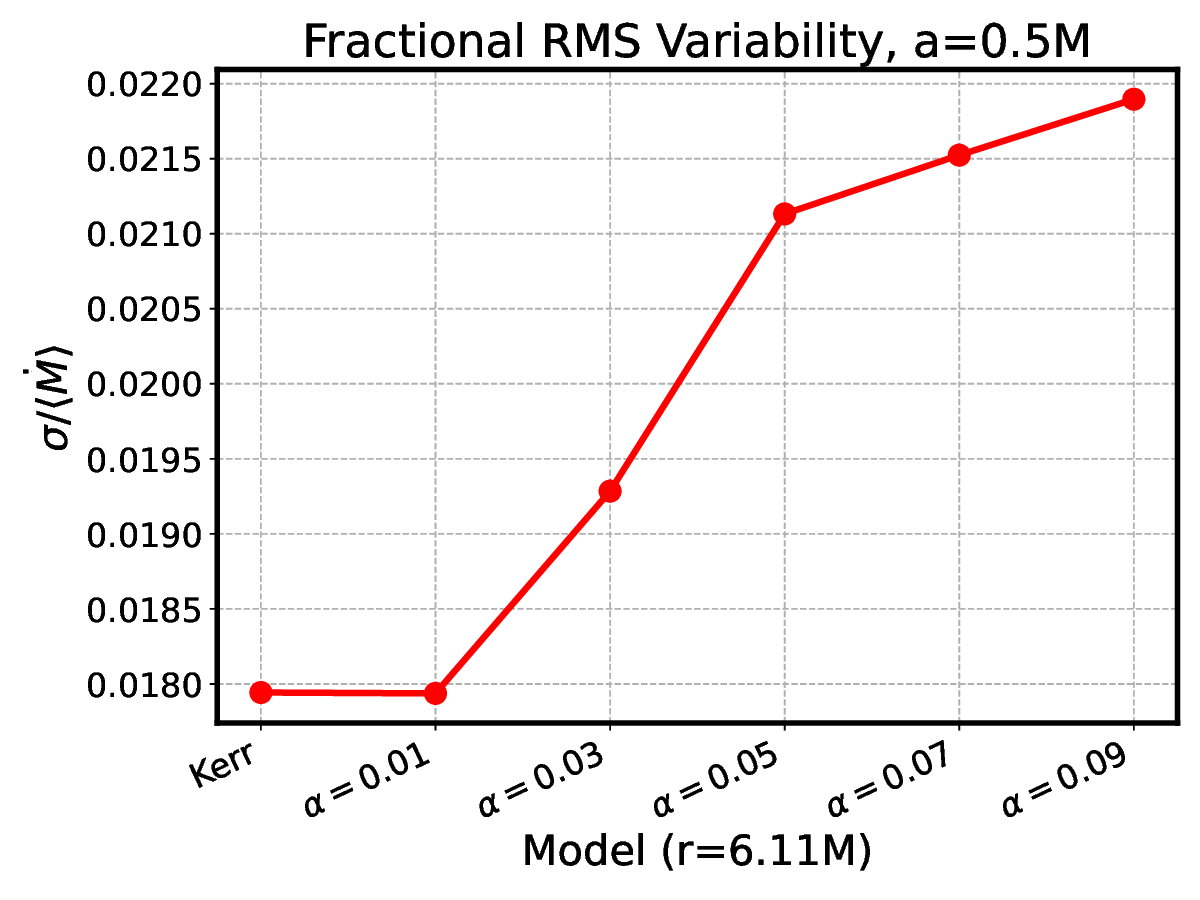,width=5.5cm,height=6.0cm} 
   \psfig{file=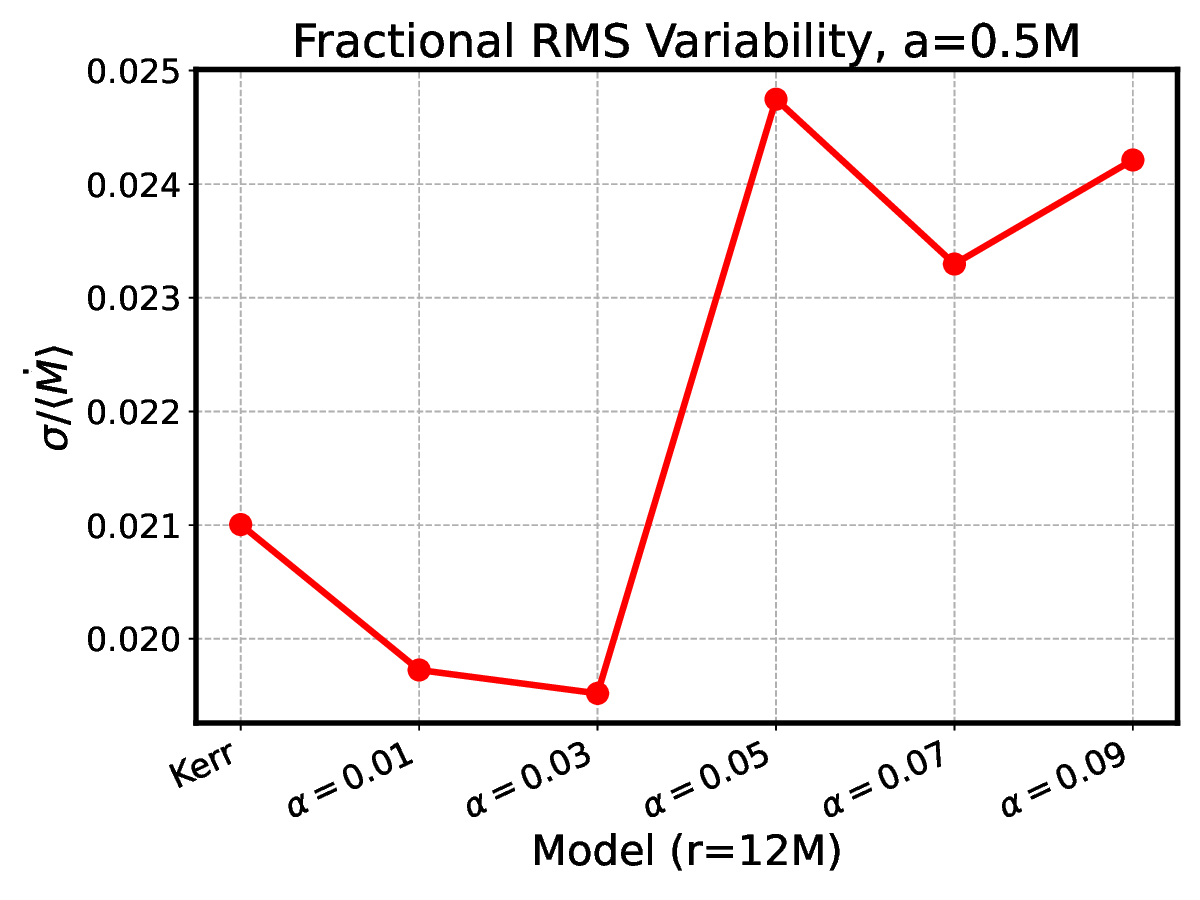,width=5.5cm,height=6.0cm} 
\caption{ The statistical characterization of accretion variability for $a=0.5M$ is presented, where the upper panels display the corresponding variations in the mean accretion rate on the embedding parameter $\alpha$ at different radial locations, while the lower panels show the dependence of the fractional root-mean-square (RMS) variability. Increasing $\alpha$ enhances the fluctuations in the mass accretion rate and systematically modifies the average mass inflow.  
}
\label{mean_frac}
\end{figure*}

 In Figs.\ref {mass_acc_a1} and \ref {mass_acc_A1B1}, the mass accretion rate is computed at different radial locations ($r = 2.3M$, $r = 6.11M$, and $r = 12M$) for the models listed in Table \ref {ST_param}, in order to demonstrate how the electromagnetic radiation that may be observed in such systems can be distinguished in terms of its dependence on the black hole spin parameter $a$, the embedding parameter $\alpha$, and the region where it is produced. The upper panels of Fig.\ref {mass_acc_a1} present the variations of the mass accretion rate around embedded black holes with rotation parameter $a = 0.5M$ for different values of $\alpha$, in comparison with the Kerr black hole model. As clearly seen, when the rotation parameter is moderate, strong QPOs arise at all radial locations around both Kerr and embedded black holes. These oscillations indicate that QPO formation is a very natural outcome in such systems and that the resulting behavior is comparable with observational features. As is theoretically known and numerically demonstrated in the previous section, the embedding parameter weakens the gravitational attraction in the strong gravitational-field region and consequently reduces the strength of the frame-dragging effect. Therefore, in regions where frame dragging is expected to be strongest, the weakening of frame dragging in embedded black hole spacetimes leads to a larger amount of matter falling toward the black hole when compared with the Kerr case. This behavior is clearly illustrated in the upper panels of Fig.\ref {mass_acc_a1}  at $r = 2.3M$. As the embedding parameter increases, the weakening of the frame-dragging effect becomes more pronounced, resulting in a larger amount of infalling matter and an enhanced mass accretion rate. At the same time, the oscillations of the mass accretion rate in embedded cases are significantly stronger than those observed in the Kerr spacetime. On the other hand, at $r = 6.11M$ and $r = 12M$, where the frame-dragging effect is either very weak or nearly absent, the mass accretion rate in embedded cases is observed to be slightly lower than that of the Kerr model as the embedding parameter increases. Nevertheless, the amplitude of the oscillations remains much stronger in the embedded cases. A more detailed discussion of the oscillation amplitudes is deferred to Fig.\ref{mean_frac}. These results indicate that the influence of the embedding parameter extends to larger spatial scales through its effect on the global shock-cone structure.

 In the lower panels of Fig.\ref {mass_acc_a1}, the Kerr case is compared with the embedded configuration A1 for a rapidly rotating black hole with rotation parameter $a = 0.9M$. When the black hole rotates rapidly, the amplitude of the mass accretion rate oscillations in the strong gravitational-field region is suppressed by the frame-dragging effect \cite{2012MNRAS.426.1533D,Donmez:2024luc,Mustafa:2025mkc,Donmez:2025piv}. As the distance from the black hole increases, the oscillation amplitude increases significantly, since the influence of frame dragging decreases exponentially with distance from the black hole horizon. This behavior is clearly observed in the lower panels of Fig.\ref {mass_acc_a1} for both the Kerr and embedded black hole models with rapid rotation. At $r = 2.3M$, the oscillation amplitudes are very small, whereas at $r = 6.11M$ and $r = 12M$ the amplitudes increase significantly. On the other hand, when the mass accretion rates around Kerr and embedded black holes are compared at $r = 2.3M$, the embedded case exhibits a relatively larger oscillation amplitude. This is due to the weakening of the frame-dragging effect caused by the embedding parameter $\alpha = 0.01M^{2}$, which leads to an increase in the oscillation amplitude of the mass accretion rate.  By contrast, at $r = 6.11M$ and $r = 12M$, both the Kerr and embedded black hole models exhibit large-amplitude oscillations with significant variability. However, the embedded black hole continues to show stronger and more irregular oscillations, suggesting that the interaction between rapid rotation and embedding corrections enhances long-term variability in the accretion flow.

Fig.\ref {mass_acc_A1B1} clearly illustrates the qualitative and quantitative behavior of the mass accretion rate at $r = 2.3M$ in the strong gravitational-field region for rapidly rotating ($a = 0.9M$) and moderately rotating ($a = 0.5M$) black hole cases, for both Kerr and embedded black hole configurations with the same embedding parameter $\alpha = 0.01M^{2}$. In the rapidly rotating black hole case, the mass accretion rate exhibits a relatively smoother temporal evolution compared to the moderate rotation case, and the oscillation amplitudes are significantly suppressed in both the Kerr and embedded configurations. This behavior indicates that strong frame-dragging effects in the strong gravitational-field region stabilize the accretion flow and damp short-timescale oscillations. Even for a small value of the embedding parameter $\alpha = 0.01M^{2}$, the mass accretion rate in the embedded configuration is systematically higher than that of the Kerr case. This demonstrates that embedding effects modify the effective gravitational potential and weaken the stabilizing influence of frame dragging. In contrast, for the moderately rotating black hole with $a = 0.5M$, although the mass accretion rate is evaluated at the same radial location $r = 2.3M$, significantly stronger and more irregular oscillations are observed. This indicates that as the frame-dragging effect weakens, hydrodynamic instabilities develop more efficiently, leading to enhanced oscillatory behavior and potentially stronger QPO signatures. The embedded configuration again exhibits stronger variability than the Kerr model, demonstrating that embedding effects amplify dynamical activity even when the rotation is moderate.

The combined presentation of the mean mass accretion rate and the fractional RMS variability provides insight into the physical characteristics of accretion dynamics in embedded black hole geometries. While the mean accretion rate reflects the efficiency of mass inflow toward the black hole, the fractional RMS variability directly indicates the degree of dynamical instability, turbulence, and oscillatory activity in the flow. When considered together, these quantities reveal whether embedding effects primarily modify the average gravitational capture of matter or amplify time-dependent fluctuations around the mean flow. This distinction is crucial for understanding how accretion variability translates into observable luminosity modulations and QPO signatures.

In the upper panel of Fig.\ref{mean_frac}, the mean accretion rate responds to embedding effects in a strongly radius-dependent manner. In the strong gravitational-field region at $r = 2.3M$, the mean accretion rate is observed to increase with increasing values of $\alpha$. As discussed previously, this behavior indicates that the stabilizing influence of frame dragging weakens as $\alpha$ increases. At the same time, it confirms that a larger amount of matter falls toward the black hole. The simultaneous increase in the amplitude of variability shows that the enhanced mass inflow is intrinsically unsteady, exhibiting stronger turbulent behavior and forming a dynamically more unstable accretion environment.In contrast, at $r = 6.11M$ and $r = 12M$, the mass accretion rate decreases gradually with increasing $\alpha$, although the variability remains strong. This radial behavior indicates that embedding effects lead to a redistribution of matter within the flow, cause an increase in the opening angle of the shock cone, and result in a reduced net inflow at larger distances. Nevertheless, the amplitude of the oscillations remains large at these radial locations.

In the lower panel of Fig.\ref{mean_frac}, the fractional RMS variability increases systematically at all radii as the embedding parameter $\alpha$ increases. In the strong gravitational-field region at $r = 2.3M$, the monotonic increase with $\alpha$ arises because the modification of the effective gravitational potential destabilizes the shock cone structure in regions close to the horizon. As the frame-dragging effect is weakened by the embedding correction, the influence of the black hole spin parameter is correspondingly reduced. This allows hydrodynamic perturbations and oscillation modes to grow more efficiently. Consequently, this region represents the most important zone where strong QPO-like variations are generated and where embedded spacetime effects can be tested. Around the ISCO, at $r = 6.11M$, a behavior similar to that observed at $r = 2.3M$ is present. This indicates that although the direct influence of frame dragging diminishes with increasing radius, embedding effects continue to reshape the global shock cone geometry, leading to enhanced variability at distances where the flow is less tightly bound but remains dynamically active. At larger distances in the strong gravitational field regime, such as $r = 12M$, the RMS variability remains high and continues to show an increasing trend with $\alpha$. This demonstrates that oscillations generated within the shock cone are not confined to the inner region but instead propagate outward. As a result, long-term variability develops throughout the accretion flow, leading to the conclusion that embedding effects influence the accretion axis globally rather than being restricted to the vicinity of $r = 2.3M$.

\subsection{QPO Imprints of Embedding Effects in BHL Accretion Flows and Their Observability}\label{S4_3} 

\begin{figure*}
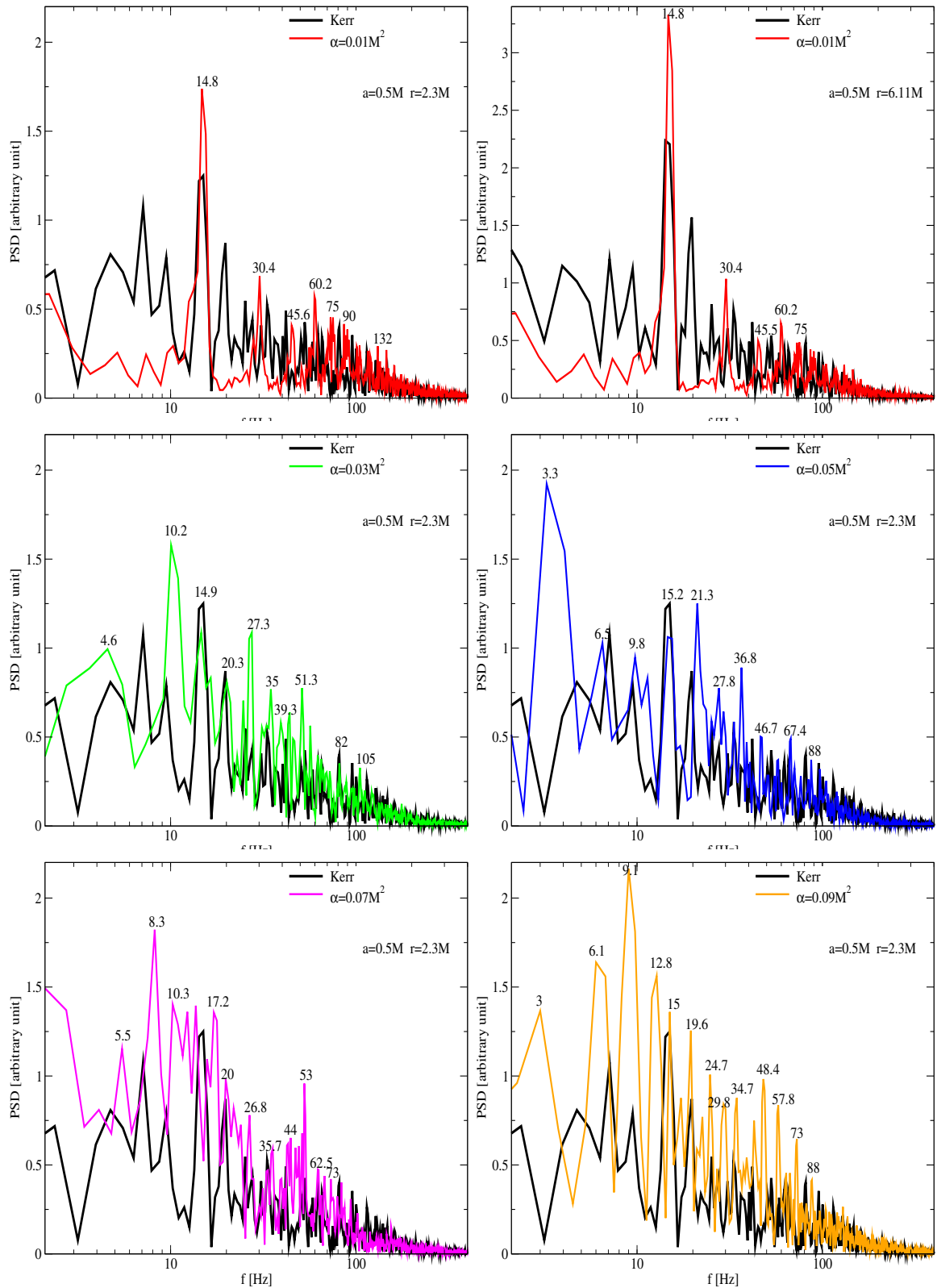

  \center
   \psfig{file=PSD_r23_B1.eps,width=7.5cm,height=7.0cm} 
   \psfig{file=PSD_r611_B1.eps,width=7.5cm,height=7.0cm} \\   
   \psfig{file=PSD_r23_B2.eps,width=7.5cm,height=7.0cm}  
   \psfig{file=PSD_r23_B3.eps,width=7.5cm,height=7.0cm}  \\ 
   \psfig{file=PSD_r23_B4.eps,width=7.5cm,height=7.0cm}  
   \psfig{file=PSD_r23_B5.eps,width=7.5cm,height=7.0cm}     
\caption{ The power spectral density (PSD) analysis computed from the mass accretion rate is presented for embedded black hole models (B1-B5) with rotation parameter $a=0.5M$ in comparison with the Kerr model, where the PSD in each case is calculated from the mass accretion rate at $r=2.3M$, while the upper-right panel shows the corresponding PSD analysis performed at $r=6.11M$ using the same parameters upper-left one, demonstrating that the resulting peaks arise globally and are independent of the radial location. The emergence of multiple peaks and frequency shifts reveals the imprint of embedding effects on QPO-like features in the accretion flow.  
}
\label{QPOs_a05}
\end{figure*}

Since the mass accretion rates exhibit strong instabilities at different radial locations, and since this behavior changes depending on the embedding black hole parameter $\alpha$, this raises the possibility that QPO formation and observational results can be explained through these numerical QPOs. For this reason, in this section we perform a PSD analysis. As already discussed, our numerical modeling results have shown that the dynamical structure of the shock cone changes as a function of the parameter $\alpha$. This variation implies that QPO frequencies dependent on the embedding parameter $\alpha$ may emerge. Motivated by this point, in Fig.\ref{QPOs_a05} we first present the PSD analysis results obtained around moderately rotating black holes. As seen in Fig.\ref{QPOs_a05}, for each panel and for a given value of the embedding parameter $\alpha$, the QPO frequencies trapped within the shock cone around embedded black holes are revealed, while in the same panel the QPO peaks produced in the Kerr case are also plotted, allowing for a direct comparison between the embedded and Kerr models.

In the upper two panels of Fig.\ref{QPOs_a05}, the PSD analyses for the embedded black hole model with $\alpha = 0.01M^{2}$ are presented in comparison with the Kerr case at two different radial locations, $r = 2.3M$ and $r = 6.11M$. The characteristic frequencies obtained at both radial positions are identical. This demonstrates that the nature of the oscillation modes is global, indicating that they are not produced by numerical artifacts or local turbulence. Instead, the oscillation modes observed in the PSD analysis arise from large-scale dynamical processes in the accretion flow, excited by the shock cone. The rich peak structure emerges as a result of nonlinear couplings among the fundamental modes. The fact that the frequencies of the peaks are independent of the radial location strongly reinforces the physical nature of these modes. Consequently, these modes appear as potential sources for observational signatures.

Another important result obtained from the upper two panels of Fig.\ref{QPOs_a05} is that the amplitudes of the peaks formed at these two radial locations, particularly for the fundamental frequency at $14.8\,\mathrm{Hz}$, are different. In the PSD analysis at $r = 2.3M$, the peak observed in the strong gravitational-field region is suppressed due to the influence of frame dragging, and as a result the oscillation amplitude is reduced. Consequently, the amplitude of this frequency at $r = 2.3M$ is lower than that at $r = 6.11M$. This result is fully consistent with the discussion above, which indicates that strong frame-dragging effects in the strong gravitational-field region act to reduce the oscillation amplitudes.

The PSD analyses shown in the middle and bottom panels of Fig.\ref{QPOs_a05} present, in comparison with the Kerr case, the oscillation modes formed around embedded black holes for increasing values of the embedding parameter $\alpha$, together with their rich spectral structure. As $\alpha$ increases, the PSD becomes progressively more complex and leads to the formation of a larger number of peaks. It is observed that the peaks produced for increasing $\alpha$ shift toward the low-frequency QPO (LFQPO) regime. At the same time, the spectral complexity also increases. This indicates that embedding effects give rise to several new persistent modes in the accretion flow that are either absent or only weakly present in the Kerr case.

In addition to the effects of the embedding parameter discussed above, it also leads to shifts in the characteristic frequencies. This variation is directly correlated with changes in the structure of the shock cone. In this context, these modes can generally be interpreted as fundamental modes trapped within the cavity formed inside the shock cone and excited through their nonlinear couplings. As a result, $\alpha$ does not merely amplify variability but actively reshapes the frequency content of the accretion flow. When all models are compared with the Kerr case, the strongest deviation is observed for the embedded configuration with $\alpha = 0.09M^{2}$. In this case, the PSD analysis shows a significant enhancement of LFQPOs together with the formation of a large number of peaks.

When the observationally detected QPO frequencies are taken into account, these results have very important implications. The global nature of the oscillation modes indicates that the same frequencies can be observed in different emission regions around the black hole. This significantly enhances the observability of these QPO frequencies. Moreover, the increased likelihood of LFQPO features with increasing $\alpha$, together with the richness in the number of peaks in the PSD spectrum, emerges as an important indication for testing embedded black holes. Accreting black hole systems that exhibit multiple LFQPOs or unusually complex timing spectra, beyond what is expected from Kerr-based models, may therefore point toward deviations in the underlying spacetime geometry.

In the PSD analyses of the embedded black hole models shown in Fig.\ref {QPOs_a05}, it is found that the commensurate frequency ratios that emerge depend clearly and systematically on the embedding parameter $\alpha$. For the small value $\alpha = 0.01M^{2}$, the spectrum is dominated by the fundamental frequency at $14.8\,\mathrm{Hz}$, which has the largest amplitude, together with an approximately harmonic overtone sequence at $30.4\,\mathrm{Hz}$, $60.2\,\mathrm{Hz}$, and $90\,\mathrm{Hz}$, clearly forming a $2{:}1$ ratio. For example, the pairs $30.4{:}14.8$ and $60.2{:}30.4$ closely satisfy this ratio. This indicates that the oscillation dynamics in this regime are produced by simple harmonic excitations of global fundamental modes.

For the embedding parameter value $\alpha = 0.03M^{2}$ given in Fig.\ref {QPOs_a05}, corresponding to an increased embedding strength, new low-frequency modes are observed to appear, and the overtone structure becomes more complex. In this case, a clear $3{:}2$ pair is formed by the frequencies at $14.9\,\mathrm{Hz}$ and $10.2\,\mathrm{Hz}$. The emergence of the $3{:}2$ ratio at this value of $\alpha$ indicates that nonlinear coupling modes become active. These nonlinear couplings allow different global modes to interact and remain simultaneously excited within the shock-cone cavity.

For $\alpha = 0.05M^{2}$, both harmonic and non-harmonic overtones are observed. The frequency pair $6.1{:}3.3$ approximately forms a $2{:}1$ ratio, while $15.2{:}9.8$ closely satisfies a $3{:}2$ ratio. The coexistence of both ratios within the same model indicates that increasing $\alpha$ not only strengthens harmonic excitations but also enhances nonlinear interactions. Such commensurate pairs are known in the literature as frequency pairs observed from the same source~\cite{Varniere:2018zea,McClintock:2007ge}.

When the embedding parameter reaches $\alpha = 0.07M^{2}$ given in Fig.\ref {QPOs_a05}, LFQPOs become more prominent. In this regime, the pair $8.3{:}5.5$ approximately forms a $3{:}2$ ratio. On the other hand, higher-frequency peaks continue to appear in this model, and these frequencies generally form integer-multiple overtone structures. This behavior indicates that strong modifications of the shock-cone structure accelerate the transition toward LFQPO-dominated dynamics.

Finally, for the strongest embedding case considered in Fig.\ref {QPOs_a05}, $\alpha = 0.09M^{2}$, the PSD analysis clearly reveals a rich overtone structure. Many low-frequency peaks are produced, together with peaks that can be classified as moderate- and high-frequency components. The pairs $6.1{:}3.0$, which closely satisfy a $2{:}1$ ratio, and $24.7{:}12.8$, which approximately form a $3{:}2$ ratio, represent observable overtones in this strong embedding regime. In addition to these, other observable peak pairs are also present. This indicates that nonlinear coupling remains active and that the accretion flow supports a broad hierarchy of interacting with global modes.

\begin{figure*}
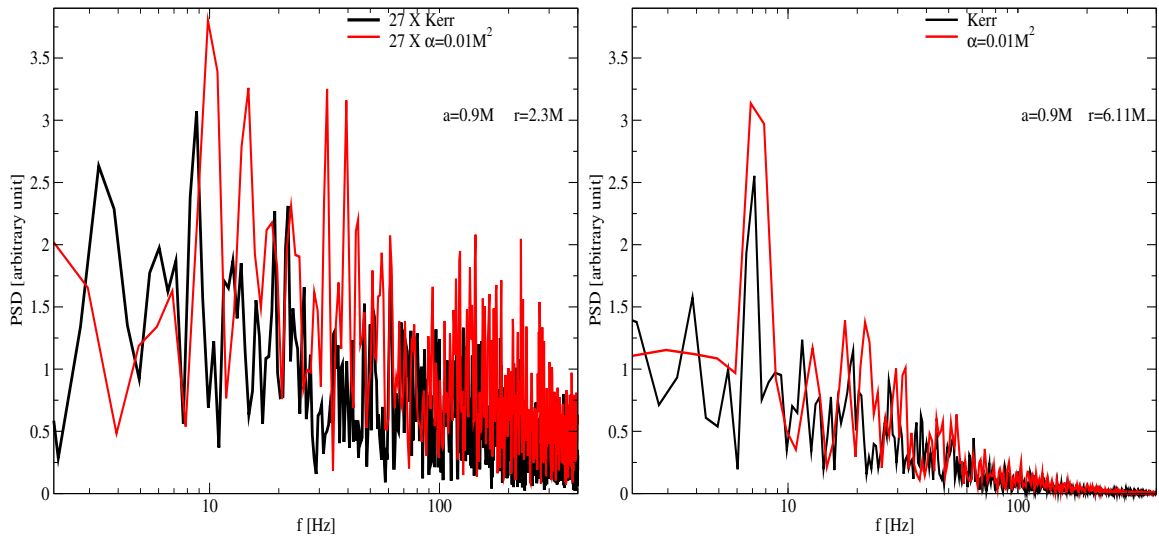

  \center
   \psfig{file=PSD_a09_r23.eps,width=7.5cm,height=7.0cm} 
   \psfig{file=PSD_a09_r611.eps,width=7.5cm,height=7.0cm} \\   
\caption{ As in Fig.\ref {QPOs_a05}, the PSD analysis is performed for the case $a=0.9M$ at different radial locations $r=2.3M$ and $r=6.11M$, where the embedding parameter $\alpha$ causes shifts in the resulting QPO frequencies, while on the other hand the rapidly rotating black hole strongly suppresses the amplitudes of the peaks formed in the strong gravitational-field region. 
}
\label{QPOs_a09}
\end{figure*}

As seen in Fig.\ref {QPOs_a09}, in the strong gravitational-field region at $r = 2.3M$, which is the closest location to the black hole horizon, the rapid rotation of the black hole with $a = 0.9M$ leads to a very strong frame-dragging effect. As a result, strong unstable oscillations are significantly stabilized and their amplitudes are reduced. This behavior is clearly demonstrated in Fig.\ref {QPOs_a09}. The PSD peaks computed from the mass accretion rate at $r = 2.3M$ have very small amplitudes, being nearly a factor of $27$ lower than those at $r = 6.11M$, while the peaks at $r = 6.11M$ are comparable in amplitude to those obtained in the PSD analyses shown in Fig.\ref {QPOs_a05}. At the same time, when the embedded black hole model for the rapidly rotating case is considered in Fig.\ref {QPOs_a09}, it becomes evident that the embedding parameter $\alpha$ acts jointly with the black hole spin $a$. As discussed in detail earlier, the embedding parameter modifies the gravitational potential and weakens the effect of frame dragging. As a consequence, the structure of the shock cone is altered. In the rapidly rotating black hole model, a nonlinear interaction between $a$ and $\alpha$ is observed, particularly in the region close to the black hole horizon. At $r = 2.3M$, where both rotation and embedding corrections are strongest, this coupling produces substantial shifts in the characteristic frequencies and modifies the overall PSD structure. This, in turn, leads to the formation of completely different frequencies at $r = 6.11M$, where the effects of these parameters are relatively weaker. Consequently, at this radius the PSD peaks depend only weakly on the interaction between $a$ and $\alpha$. As a result, different and generally more coherent frequency peaks are formed.

\section{Conclusions}\label{S5} 

The analysis of particle motion around a rotating and embedded BH demonstrates systematic and quantifiable effects of the rotation parameter ($a$) and the embedding parameter ($\alpha$) on orbital characteristics and stability. In the general configuration where both $a$ and $\alpha$ are nonzero, the conserved energy ($\mathcal{E}$) and angular momentum ($\mathcal{L}$) decrease as either parameter increases, indicating stronger gravitational binding resulting from the combined influence of frame-dragging and embedding-induced curvature modifications. In this case, the effective potential exhibits shallower minima and a reduced radial range, showing that stable circular orbits are confined closer to the BH and that the ISCO's radius is smaller compared to standard Kerr or Schwarzschild geometries. Also, the corresponding effective radial force increases with $a$ and $\alpha$, reflecting the amplified gravitational attraction induced by both rotation and embedding. In the limit $\alpha \to 0$ with finite $a$, the metric reduces to the Kerr solution, restoring the standard frame-dragging features while removing embedding contributions,  energy and angular momentum increase relative to the embedded case, the effective potential becomes deeper, and circular orbits are less tightly bound, consistent with weaker overall gravitational enhancement. Conversely, for $a \to 0$ with finite $\alpha$, the spacetime corresponds to a static embedded BH, where rotational effects are absent but embedding corrections continue to reduce $\mathcal{E}$ and $\mathcal{L}$, lower the potential depth, and increase the effective force relative to Schwarzschild, demonstrating the purely geometric influence of embedding on particle dynamics. In the double limit $a \to 0$ and $\alpha \to 0$, the spacetime reduces to Schwarzschild, yielding the largest energy and angular momentum for circular orbits, the deepest effective potential well, and the weakest gravitational pull, consistent with the canonical description of stable circular geodesics in GR. Also, comparison of these limiting cases indicates that rotation and embedding similarly decrease energy and angular momentum, strengthen gravitational binding, and reduce orbital stability, but their physical origins differ: $a$ generates frame-dragging and relativistic precessions, whereas $\alpha$ alters the spacetime curvature through embedding effects. In this context, the combined variation of $a$ and $\alpha$ thus defines a continuous framework connecting Schwarzschild, Kerr, and embedded geometries, allowing a quantitative assessment of the separate contributions of rotation and embedding to orbital dynamics, effective potential structure, force magnitude, and precessional behavior. 

The analysis of small harmonic perturbations around circular orbits of neutral test particles in a rotating and embedded BH spacetime exhibits a detailed structure of epicyclic motion determined by the radial, vertical, and azimuthal frequencies. In this case, locally measured frequencies $(\omega_r, \omega_\theta, \omega_\phi)$ quantify the particle response to radial and vertical perturbations as well as its angular motion, while the corresponding redshifted frequencies $(\Omega_r, \Omega_\theta, \Omega_\phi)$, as observed at spatial infinity, incorporate gravitational time dilation and frame-dragging effects. In the static limit $(a=0)$, the radial and vertical frequencies remain degenerate due to the spacetime's spherical symmetry, and the embedding parameter $(\alpha)$ induces a moderate inward shift of frequency extrema, introducing additional relativistic corrections without breaking the degeneracy. When the embedding parameter vanishes $(\alpha=0)$, the spacetime reduces to the Kerr solution, where rotation alone lifts the degeneracy between radial and vertical modes and shifts the extrema to smaller radii closer to the horizon. In the combined limit $(a=0, \alpha=0)$, corresponding to the Schwarzschild case, all three frequencies converge in the Newtonian regime at large radii, whereas near the BH the radial frequency becomes smaller than the vertical and azimuthal frequencies, generating relativistic periapsis precession. Also, the periapsis precession frequency $(\Omega_P = \Omega_\phi - \Omega_r)$ systematically decreases with increasing spin or embedding, reflecting a reduction in precessional effects from frame dragging or spacetime embedding. In contrast, the Lense-Thirring precession $(\Omega_{LT})$ increases with BH rotation, consistent with inertial frame dragging, but is attenuated by the embedding parameter, which counteracts the rotational contribution. In this context, these results illustrate that the spin parameter $(a)$ primarily controls the separation of epicyclic modes and the magnitude of frame-dragging precession, while the embedding parameter $(\alpha)$ predominantly adjusts the radial positions of extrema and contributes additional relativistic modifications. Also, the analysis demonstrates that the relation between $a$ and $\alpha$ consistently alters local and observable frequencies, periapsis precession, and Lense-Thirring effects, and all relevant limits reproduce classical and relativistic behaviors, establishing a coherent framework for test particle dynamics in generalized BH spacetimes. 

In the second part of the paper, we numerically solved the GRH equations in order to analyze the embedding-parameter-dependent variations in the dynamical structure of the shock cone formed via BHL accretion around the rotating embedded black holes. By comparing different embedding configurations with the Kerr case, we aimed to understand the changes in shock cone morphology, the behavior of the accretion dynamics, and the resulting observable signatures. Through a systematic variation of the black hole spin parameter $a$ and the embedding parameter $\alpha$, we revealed parameter-dependent modifications in both strong and relatively weak gravitational-field regions. The numerical results show that the BHL mechanism is a crucial process for unveiling the effects of strong gravity through the structure of the formed shock cone. With increasing values of the embedding parameter $\alpha$, we demonstrated a widening of the shock-cone opening angle, a suppression of the density of matter trapped inside the cone, a weakening of the post-shock compression, and a redistribution of matter within the cone. While these effects lead to strong oscillations in the mass accretion rate, we also numerically showed that the cone captures and excites possible fundamental modes around the black hole. These morphological changes directly impact the time-dependent accretion behavior and enhance the overall variability of the flow compared to the Kerr case.

On the other hand, the time evolution of the mass accretion rate shows that embedding effects systematically enhance the dynamical activity in the accretion flow. In the strong gravitational-field region, the frame-dragging effect of the rotating black hole is observed to stabilize the instabilities developing in the flow and to reduce the amplitudes of time-dependent oscillations. However, since the embedding parameter weakens the frame-dragging effect, this stabilization is observed to decrease as the embedding parameter increases. Because the influence of the black hole spin parameter is significantly reduced in the vicinity of the ISCO, while the embedding parameter still retains a non-negligible effect in this region, the instabilities observed in the mass accretion rate around the ISCO are found to be stronger and more pronounced than those in the deep strong gravitational-field region.

 One of the most important outcomes of numerical simulations is the PSD analysis. The PSD analyses help us to understand the QPOs generated by strong oscillations developing around black holes. In this context, by using the mass accretion rates, we performed PSD analyses for each model and revealed how the amplitudes of the resulting peaks and their corresponding frequencies vary depending on the parameters $\alpha$ and $a$. The numerical results show that with increasing values of $\alpha$, a systematic enrichment of the frequency spectrum is observed. At the same time, the number of LFQPOs increases with increasing $\alpha$, and these features appear as more pronounced and observable peaks. Furthermore, the observability of commensurate frequency ratios such as $3{:}2$ and $2{:}1$ is found to increase with increasing $\alpha$. The frequencies forming these ratios arise entirely from the excitation of fundamental modes trapped within the shock cone. Since the parameters $a$ and $\alpha$ modify the structure of the shock cone, the resulting QPO frequencies, the amplitudes of the peaks, and their agreement with observations are found to depend sensitively on the model parameters. In addition, particularly for the rapidly rotating black hole models with $a = 0.9M$, the strong coupling between $a$ and $\alpha$ leads to variations in the frequencies measured at different radial locations in the PSD analysis.

Lastly,  this study shows the consistency between the theoretical results derived from test-particle motion around rotating embedded black holes and the numerical results obtained from solving the GRH equations for the BHL accretion mechanism. In theoretical studies, it has been shown that with increasing values of $\alpha$, relativistic precession effects weaken, as inferred from the behavior of epicyclic, periapsis precession, and Lense-Thirring frequencies. The numerical results obtained from hydrodynamic solutions confirm this behavior. The suppression of the frame-dragging effect by the embedding constant $\alpha$ leads to systematic shifts in the characteristic QPO frequencies that emerge. At the same time, a reduction in precessional effects is observed in the strong gravitational-field region. Moreover, the commensurate frequency ratios revealed by the PSD analyses are consistent with the behavior of coupled oscillation modes influenced by the same underlying spacetime modifications. This agreement between test-particle theory and full GRHD simulations strengthens the interpretation that the observed QPO features originate from genuine spacetime effects rather than purely hydrodynamic artifacts, reinforcing the role of BHL accretion flows as sensitive probes of black hole geometry.

\section*{Acknowledgments}
All numerical simulations were performed using the Phoenix High
Performance Computing facility at the American University of the Middle East (AUM), Kuwait.

\section*{Data Availability Statement}
The data sets generated and analyzed during the current study were produced using high-performance computing resources. These data are not publicly available due to their large size and computational nature, but are available from the corresponding author upon reasonable request.


\bibliography{bibwithouttitle}

@article{Donmez:2013qxa,
    author = "Donmez, Orhan",
    title = "{On the development of the Papaloizou{\textendash}Pringle instability of the black hole{\textendash}torus systems and quasi-periodic oscillations}",
    eprint = "1304.0584",
    archivePrefix = "arXiv",
    primaryClass = "astro-ph.HE",
    doi = "10.1093/mnras/stt2255",
    journal = "Mon. Not. Roy. Astron. Soc.",
    volume = "438",
    number = "1",
    pages = "846--858",
    year = "2014"
}

@article{Donmez:2017gdp,
    author = "Donmez, Orhan",
    title = "{Angular velocity perturbations inducing the Papaloizou{\textendash}Pringle instability and QPOs in the torus around the black hole}",
    eprint = "1703.08175",
    archivePrefix = "arXiv",
    primaryClass = "astro-ph.HE",
    doi = "10.1142/S0217732317501085",
    journal = "Mod. Phys. Lett. A",
    volume = "32",
    number = "20",
    pages = "1750108",
    year = "2017"
}

@ARTICLE{2012MNRAS.426.1533D,
       author = {{D{\"o}nmez}, O.},
        title = "{Relativistic simulation of flip-flop instabilities of Bondi-Hoyle accretion and quasi-periodic oscillations}",
      journal = {MNRAS},
         year = 2012,
        month = oct,
       volume = {426},
       number = {2},
        pages = {1533-1545},
          doi = {10.1111/j.1365-2966.2012.21616.x}
}

@article{Donmez:2024luc,
    author = "Donmez, Orhan and Dogan, Fatih",
    title = "{Estimating the possible QPOs of M87∗ from the parameters of a hairy Kerr black hole}",
    eprint = "2407.01478",
    archivePrefix = "arXiv",
    primaryClass = "gr-qc",
    doi = "10.1016/j.dark.2024.101718",
    journal = "Phys. Dark Univ.",
    volume = "46",
    pages = "101718",
    year = "2024"
}

@article{Mustafa:2025mkc,
    author = "Mustafa, G. and Ghosh, Sushant G. and Donmez, Orhan and Maurya, S. K. and Orzuev, Shakhzod and Atamurotov, Farruh",
    title = "{Testing Quantum-Corrected Black Holes with QPOs Observations: A Study of Particle Dynamics and Accretion Flow}",
    eprint = "2506.16405",
    archivePrefix = "arXiv",
    primaryClass = "gr-qc",
    doi = "10.1088/1475-7516/2025/10/068",
    journal = "JCAP",
    volume = "10",
    pages = "068",
    year = "2025"
}

@article{Donmez:2025piv,
    author = "Donmez, O.",
    title = "{Accretion dynamics and QPO signatures around quantum-corrected black hole: a comparison with Kerr spacetime}",
    doi = "10.1140/epjc/s10052-025-14779-6",
    journal = "Eur. Phys. J. C",
    volume = "85",
    number = "9",
    pages = "1019",
    year = "2025"
}

@article{Varniere:2018zea,
    author = "Varniere, Peggy and Rodriguez, Jerome",
    title = "{Looking for the Elusive 3:2 Ratio of High-frequency Quasi-periodic Oscillations in the Microquasar XTE J1550{\ensuremath{-}}564}",
    eprint = "1808.06823",
    archivePrefix = "arXiv",
    primaryClass = "astro-ph.HE",
    doi = "10.3847/1538-4357/aad774",
    journal = "Astrophys. J.",
    volume = "865",
    number = "2",
    pages = "113",
    year = "2018"
}

@article{McClintock:2007ge,
    author = "McClintock, Jeffrey E. and Remillard, Ronald A. and Rupen, Michael P. and Torres, M. A. P. and Steeghs, D. and Levine, Alan M. and Orosz, Jerome A.",
    title = "{The 2003 Outburst of the X-ray Transient H 1743-322: Comparisons with the Black Hole Microquasar XTE J1550-564}",
    eprint = "0705.1034",
    archivePrefix = "arXiv",
    primaryClass = "astro-ph",
    doi = "10.1088/0004-637X/698/2/1398",
    journal = "Astrophys. J.",
    volume = "698",
    pages = "1398--1421",
    year = "2009"
}

@article{Koyuncu:2014nga,
    author = {Koyuncu, Fahrettin and D{\"o}nmez, Orhan},
    title = "{Numerical simulation of the disk dynamics around the black hole: Bondi Hoyle accretion}",
    doi = "10.1142/S0217732314501156",
    journal = "Mod. Phys. Lett. A",
    volume = "29",
    pages = "1450115",
    year = "2014"
}

@article{Donmez:2022dze,
    author = "Donmez, Orhan and Dogan, Fatih and Sahin, Tugba",
    title = "{Study of Asymptotic Velocity in the Bondi{\textendash}Hoyle Accretion Flows in the Domain of Kerr and 4-D Einstein{\textendash}Gauss{\textendash}Bonnet Gravities}",
    eprint = "2205.14382",
    archivePrefix = "arXiv",
    primaryClass = "astro-ph.HE",
    doi = "10.3390/universe8090458",
    journal = "Universe",
    volume = "8",
    number = "9",
    pages = "458",
    year = "2022"
}

@article{Donmez:2026yvm,
    author = "Donmez, Orhan and Murodov, Sardor and Rayimbaev, Javlon",
    title = "{Testing strong gravitational field using the Johannsen{\textendash}Psaltis metric: Bondi{\textendash}Hoyle{\textendash}Lyttleton accretion model and QPO studies}",
    doi = "10.1016/j.aop.2026.170350",
    journal = "Annals Phys.",
    volume = "486",
    pages = "170350",
    year = "2026"
}

@article{BHP1,
  author  = {Abbott, B. P. and others},
  journal = {Phys. Rev. Lett.},
  volume  = {116},
  number  = {6},
  pages   = {061102},
  year    = {2016}
}

@article{BHP2,
  author  = {Abbott, B. P. and others},
  journal = {Phys. Rev. X},
  volume  = {9},
  number  = {3},
  pages   = {031040},
  year    = {2019}
}

@article{BHP3,
  author  = {Abbott, B. P. and others},
  journal = {Astrophys. J. Lett.},
  volume  = {892},
  number  = {1},
  pages   = {L3},
  year    = {2020}
}

@article{BHP4,
  author  = {Akiyama, K. and others},
  journal = {Astrophys. J. Lett.},
  volume  = {875},
  pages   = {L1},
  year    = {2019}
}

@article{BHP5,
  author  = {Akiyama, K. and others},
  journal = {Astrophys. J. Lett.},
  volume  = {875},
  number  = {1},
  pages   = {L6},
  year    = {2019}
}

@article{BHP6,
  author  = {Akiyama, K. and others},
  journal = {Astrophys. J. Lett.},
  volume  = {875},
  number  = {1},
  pages   = {L4},
  year    = {2019}
}

@article{BHP7,
  author  = {Akiyama, K. and others},
  journal = {Astrophys. J. Lett.},
  volume  = {930},
  number  = {2},
  pages   = {L12},
  year    = {2022}
}

@article{BHP8,
  author  = {Akiyama, K. and others},
  journal = {Astrophys. J. Lett.},
  volume  = {930},
  number  = {2},
  pages   = {L17},
  year    = {2022}
}

@inproceedings{BHP9,
  author    = {Bardeen, J. M.},
  booktitle = {Proceedings of the International Conference on Relativity Theory and Astrophysics (GR5)},
  year      = {1968}
}

@article{BHP10,
  author  = {Ay{\'o}n-Beato, E. and Garc{\'\i}a, A.},
  journal = {Phys. Lett. B},
  volume  = {493},
  pages   = {149},
  year    = {2000}
}

@article{BHP11,
  author  = {Dymnikova, I.},
  journal = {Gen. Relativ. Gravit.},
  volume  = {24},
  pages   = {235},
  year    = {1992}
}

@article{BHP12,
  author  = {Nicolini, P. and Smailagic, A. and Spallucci, E.},
  journal = {Phys. Lett. B},
  volume  = {632},
  pages   = {547},
  year    = {2006}
}

@article{BHP13,
  author  = {Balakin, A. B. and Lemos, J. P. S. and Zayats, A. E.},
  journal = {Phys. Rev. D},
  volume  = {93},
  number  = {8},
  pages   = {084004},
  year    = {2016}
}

@article{BHP14,
  author  = {Roupas, Z.},
  journal = {Eur. Phys. J. C},
  volume  = {82},
  number  = {3},
  pages   = {255},
  year    = {2022}
}

@article{BHP15,
  author  = {Borde, A.},
  journal = {Phys. Rev. D},
  volume  = {55},
  pages   = {7615},
  year    = {1997}
}

@article{BHP16,
  author  = {Bonanno, A. and Reuter, M.},
  journal = {Phys. Rev. D},
  volume  = {62},
  pages   = {043008},
  year    = {2000}
}

@article{BHP17,
  author  = {Gambini, R. and Pullin, J.},
  journal = {Phys. Rev. Lett.},
  volume  = {101},
  pages   = {161301},
  year    = {2008}
}

@article{BHP18,
  author  = {Perez, A.},
  journal = {Rep. Prog. Phys.},
  volume  = {80},
  number  = {12},
  pages   = {126901},
  year    = {2017}
}

@article{BHP19,
  author  = {Brahma, S. and Chen, C.-Y. and Yeom, D.-H.},
  journal = {Phys. Rev. Lett.},
  volume  = {126},
  number  = {18},
  pages   = {181301},
  year    = {2021}
}

@article{BHP20,
  author  = {Torres, R.},
  journal = {arXiv},
  eprint  = {2208.12713},
  archivePrefix = {arXiv},
  primaryClass  = {gr-qc},
  year    = {2023}
}

@article{BHP21,
  author  = {Lan, C. and Yang, H. and Guo, Y. and Miao, Y.-G.},
  journal = {Int. J. Theor. Phys.},
  volume  = {62},
  number  = {9},
  pages   = {202},
  year    = {2023}
}

@article{QPOs1,
  author  = {Rayimbaev, J. and Tadjimuratov, P. and Abdujabbarov, A. and Ahmedov, B. and Khudoyberdieva, M.},
  journal = {Galaxies},
  volume  = {9},
  number  = {4},
  pages   = {75},
  year    = {2021}
}

@article{QPOs2,
  author  = {Jumaniyozov, S. and Khan, S. U. and Rayimbaev, J. and Abdujabbarov, A. and Urinbaev, S. and Murodov, S.},
  journal = {Eur. Phys. J. C},
  volume  = {84},
  number  = {9},
  pages   = {964},
  year    = {2024}
}

@article{QPOs3,
  author  = {Faraji, S.},
  journal = {Astron. Rep.},
  volume  = {67},
  pages   = {S207--S213},
  year    = {2023}
}

@article{QPOs4,
  author  = {Stuchl{\'\i}k, Z. and Vrba, J.},
  journal = {JCAP},
  volume  = {2021},
  number  = {11},
  pages   = {059},
  year    = {2021}
}

@article{QPOs5,
  author  = {Homan, J. and others},
  journal = {Astrophys. J.},
  volume  = {586},
  pages   = {1262},
  year    = {2003}
}

@article{QPOs6,
  author  = {Rayimbaev, J. and Ahmedov, B. and Bokhari, A. H.},
  journal = {Int. J. Mod. Phys. D},
  volume  = {31},
  pages   = {2240004},
  year    = {2022}
}

@article{QPOs7,
  author  = {Qi, M. and Rayimbaev, J. and Ahmedov, B.},
  journal = {Eur. Phys. J. C},
  volume  = {83},
  pages   = {730},
  year    = {2023}
}

@book{QPOs8,
  author    = {Rezzolla, L. and Zanotti, O.},
  publisher = {Oxford University Press},
  year      = {2013}
}

@article{QPOs9,
  author  = {Rezzolla, L. and Yoshida, S. and Maccarone, T. J. and Zanotti, O.},
  journal = {Mon. Not. R. Astron. Soc.},
  volume  = {344},
  pages   = {L37},
  year    = {2003}
}

@article{QPOs10,
  author  = {T{\"o}r{\"o}k, G. and Kotrlov{\'a}, A. and Sr{\'a}mkov{\'a}, E. and Stuchl{\'\i}k, Z.},
  journal = {Astron. Astrophys.},
  volume  = {531},
  pages   = {A59},
  year    = {2011}
}

@article{QPOs11,
  author  = {Stuchl{\'\i}k, Z. and Kotrlov{\'a}, A. and T{\"o}r{\"o}k, G.},
  journal = {Astron. Astrophys.},
  volume  = {525},
  pages   = {A82},
  year    = {2011}
}

@article{QPOs12,
  author  = {Stuchl{\'\i}k, Z. and Kotrlov{\'a}, A. and T{\"o}r{\"o}k, G.},
  journal = {Astron. Astrophys.},
  volume  = {552},
  pages   = {A10},
  year    = {2013}
}

@article{QPOs13,
  author  = {Mustafa, G. and Hussain, I. and Liu, W.-M.},
  journal = {Chin. J. Phys.},
  volume  = {80},
  pages   = {148},
  year    = {2022}
}

@article{QPOs14,
  author  = {Jiang, X. and Wang, P. and Yang, H. and Wu, H.},
  journal = {Eur. Phys. J. C},
  volume  = {81},
  pages   = {1043},
  year    = {2021}
}

@article{QPOs15,
  author  = {Amarilla, L. and Eiroa, E. F. and Giribet, G.},
  journal = {Phys. Rev. D},
  volume  = {81},
  pages   = {124045},
  year    = {2010}
}

@article{QPOs16,
  author  = {Liu, Y. and Mustafa, G. and Maurya, S. K. and Javed, F.},
  journal = {Eur. Phys. J. C},
  volume  = {83},
  pages   = {584},
  year    = {2023}
}

@article{QPOs17,
  author  = {Stuchl{\'\i}k, Z. and Kolo{\v s}, M.},
  journal = {Astron. Astrophys.},
  volume  = {586},
  pages   = {A130},
  year    = {2016}
}

@article{QPOs18,
  author  = {Remillard, R. A. and others},
  journal = {Astrophys. J.},
  volume  = {517},
  number  = {2},
  pages   = {L127},
  year    = {1999}
}

@article{QPOs19,
  author  = {Morgan, E. H. and Remillard, R. A. and Greiner, J.},
  journal = {Astrophys. J.},
  volume  = {482},
  number  = {2},
  pages   = {993},
  year    = {1997}
}

@article{QPOs20,
  author  = {Remillard, R. A. and others},
  journal = {Astrophys. J.},
  volume  = {637},
  number  = {2},
  pages   = {1002},
  year    = {2006}
}

@article{QPOs21,
  author  = {Shahzadi, M. and Kolo{\v s}, M. and Saleem, R. and Stuchl{\'\i}k, Z.},
  journal = {Class. Quant. Grav.},
  volume  = {41},
  number  = {7},
  pages   = {075014},
  year    = {2024}
}

@article{QPOs22,
  author  = {Morgan, E. H. and Remillard, R. A. and Greiner, J.},
  journal = {Astrophys. J.},
  volume  = {482},
  number  = {2},
  pages   = {993},
  year    = {1997}
}

@article{QPOs23,
  author  = {Jumaniyozov, S. and others},
  journal = {Eur. Phys. J. C},
  volume  = {85},
  number  = {2},
  pages   = {126},
  year    = {2025}
}

@article{QPOs24,
  author  = {Donmez, O.},
  journal = {Eur. Phys. J. C},
  volume  = {84},
  number  = {5},
  pages   = {524},
  year    = {2024}
}

@article{QPOs25,
  author  = {Garg, A. and Misra, R. and Sen, S.},
  journal = {Mon. Not. R. Astron. Soc.},
  volume  = {514},
  number  = {3},
  pages   = {3285--3293},
  year    = {2022}
}

@article{QPOs26,
  author  = {Masterson, M. and others},
  journal = {Nature},
  volume  = {638},
  pages   = {370--375},
  year    = {2025}
}

@article{QPOs27,
  author  = {Bambi, C.},
  journal = {JCAP},
  volume  = {2012},
  number  = {09},
  pages   = {014},
  year    = {2012}
}

@article{QPOs28,
  author  = {Abdulkhamidov, F. and Narzilloev, B. and Hussain, I. and Abdujabbarov, A. and Ahmedov, B.},
  journal = {Eur. Phys. J. C},
  volume  = {84},
  number  = {4},
  pages   = {420},
  year    = {2024}
}

@article{QPOs29,
  author  = {T{\"o}r{\"o}k, G.},
  journal = {Astron. Nachr.},
  volume  = {326},
  number  = {9},
  pages   = {856--860},
  year    = {2005}
}

@article{QPOs30,
  author  = {Petri, J.},
  journal = {Astrophys. Space Sci.},
  volume  = {318},
  number  = {3},
  pages   = {181--186},
  year    = {2008}
}

@article{QPOs31,
  author  = {Vignarca, F. and Migliari, S. and Belloni, T. and Psaltis, D. and Van Der Klis, M.},
  journal = {Astron. Astrophys.},
  volume  = {397},
  number  = {2},
  pages   = {729--738},
  year    = {2003}
}

@article{QPOs32,
  author  = {Pasham, D. R. and others},
  journal = {Astrophys. J. Lett.},
  volume  = {811},
  number  = {1},
  pages   = {L11},
  year    = {2015}
}

@article{QPOs33,
  author  = {Wagoner, R. V.},
  journal = {Astrophys. J. Lett.},
  volume  = {752},
  number  = {2},
  pages   = {L18},
  year    = {2012}
}

@article{QPOs34,
  author  = {Mustafa, G. and Demir, E. and Davlataliev, A. and Chaudhary, H. and Atamurotov, F. and G{\"u}dekli, E.},
  journal = {Phys. Dark Universe},
  volume  = {46},
  pages   = {101644},
  year    = {2024}
}

@article{QPOs35,
  author  = {Rayimbaev, J. and Murodov, S. and Shermatov, A. and Yusupov, A.},
  journal = {Eur. Phys. J. C},
  volume  = {84},
  number  = {10},
  pages   = {1114},
  year    = {2024}
}

@article{RefQS1,
  author  = {Bambi, C.},
  journal = {J. Cosmol. Astropart. Phys.},
  volume  = {2012},
  pages   = {014},
  year    = {2012}
}

@article{RefQS2,
  author  = {D{\"o}nmez, O.},
  journal = {Astrophys. Space Sci.},
  volume  = {293},
  pages   = {323},
  year    = {2004}
}

@article{RefQS3,
  author  = {Montero, P. J. and Zanotti, O.},
  journal = {Mon. Not. Roy. Astron. Soc.},
  volume  = {419},
  pages   = {1507},
  year    = {2011}
}

@article{RefQS4,
  author  = {Shahzadi, M. and Kolo{\v s}, M. and Saleem, R. and Stuchl{\'\i}k, Z.},
  journal = {Class. Quantum Gravity},
  volume  = {41},
  pages   = {075014},
  year    = {2024}
}

@article{RefQS5,
  author  = {Stuchl{\'\i}k, Z. and Kolo{\v s}, M. and Tursunov, A.},
  journal = {Publ. Astron. Soc. Jpn.},
  volume  = {74},
  pages   = {1220},
  year    = {2022}
}

@article{RefQS6,
  author  = {Boshkayev, K. and Luongo, O. and Muccino, M.},
  journal = {Phys. Rev. D},
  volume  = {108},
  pages   = {124034},
  year    = {2023}
}

@article{RefQS7,
  author  = {Boshkayev, K. and Konysbayev, T. and Kurmanov, Y. and Muccino, M. and Quevedo, H.},
  journal = {Mon. Not. Roy. Astron. Soc.},
  volume  = {531},
  pages   = {3876},
  year    = {2024}
}

@article{RefQS8,
  author  = {Boshkayev, K. and Idrissov, A. and Luongo, O. and Muccino, M.},
  journal = {Phys. Rev. D},
  volume  = {108},
  pages   = {044063},
  year    = {2023}
}

@article{RefQS9,
  author  = {Mitra, S. and Vrba, J. and Rayimbaev, J. and Stuchl{\'\i}k, Z. and Ahmedov, B.},
  journal = {Phys. Dark Univ.},
  volume  = {46},
  pages   = {101561},
  year    = {2024}
}

@article{RefQS10,
  author  = {Rayimbaev, J. and Eshimbetov, U. and Majeed, B. and Abdujabbarov, A. and Abduvokhidov, A. and Abdulazizov, B. and Xalmirzayev, A.},
  journal = {Chin. Phys. C},
  volume  = {48},
  pages   = {055104},
  year    = {2024}
}

@article{RefQS11,
  author  = {Abdulkhamidov, F. and Nedkova, P. and Rayimbaev, J. and Kunz, J. and Ahmedov, B.},
  journal = {Phys. Rev. D},
  volume  = {109},
  pages   = {104074},
  year    = {2024}
}

@article{RefQS12,
  author  = {Qi, M. and Rayimbaev, J. and Ahmedov, B.},
  journal = {Eur. Phys. J. C},
  volume  = {83},
  pages   = {730},
  year    = {2023}
}

@article{RefQS13,
  author  = {Rayimbaev, J. and Abdujabbarov, A. and Abdulkhamidov, F. and Khamidov, V. and Djumanov, S. and Toshov, J. and Inoyatov, S.},
  journal = {Eur. Phys. J. C},
  volume  = {82},
  pages   = {1110},
  year    = {2022}
}

@inproceedings{RefQS14,
  author    = {Stuchl{\'\i}k, Z. and Kotrlov{\'a}, A.},
  booktitle = {RAGtime 8/9: Workshops on Black Holes and Neutron Stars},
  year      = {2007}
}

@article{RefQS15,
  author  = {Stuchl{\'\i}k, Z. and Kotrlov{\'a}, A.},
  journal = {Gen. Relativ. Gravit.},
  volume  = {41},
  pages   = {1305},
  year    = {2009}
}

@article{RefQS16,
  author  = {Murodov, S. and Badalov, K. and Rayimbaev, J. and Ahmedov, B. and Stuchl{\'\i}k, Z.},
  journal = {Symmetry},
  volume  = {16},
  pages   = {109},
  year    = {2024}
}

@article{RefQS17,
  author  = {Murodov, S. and Rayimbaev, J. and Ahmedov, B. and Hakimov, A.},
  journal = {Symmetry},
  volume  = {15},
  pages   = {2084},
  year    = {2023}
}

@article{RefQS18,
  author  = {Nucamendi, U. and Becerril, R. and Sheoran, P.},
  journal = {Eur. Phys. J. C},
  volume  = {80},
  pages   = {35},
  year    = {2020}
}

@article{RefQS19,
  author  = {Aliev, A. N. and Talazan, P.},
  journal = {Phys. Rev. D},
  volume  = {80},
  pages   = {044023},
  year    = {2009}
}

@article{RefQS20,
  author  = {Jusufi, K. and Jamil, M. and Moraes, P. H. R. S.},
  journal = {Eur. Phys. J. C},
  volume  = {80},
  pages   = {354},
  year    = {2020}
}

@article{RefQS21,
  author  = {Das, S. and Vagenas, E. C.},
  journal = {Can. J. Phys.},
  volume  = {87},
  pages   = {233},
  year    = {2018}
}

@article{RefQS22,
  author  = {Jusufi, K. and Azreg-A\"{\i}nou, M. and Jamil, M. and Zhu, T.},
  journal = {Int. J. Geom. Methods Mod. Phys.},
  volume  = {19},
  pages   = {2250068},
  year    = {2022}
}

@article{RefQS23,
  author  = {Abedi, J. and Arfaei, H.},
  journal = {J. Cosmol. Astropart. Phys.},
  volume  = {2016},
  pages   = {042},
  year    = {2016}
}

@article{RefQS24,
  author  = {Banerjee, I.},
  journal = {J. Cosmol. Astropart. Phys.},
  volume  = {2022},
  pages   = {034},
  year    = {2022}
}

@misc{RefQS25,
  author  = {Motta, S. E. and Belloni, T. M.},
  howpublished = {arXiv:2307.00867},
  year    = {2023}
}

@article{RefQS26,
  author  = {Shao, C.-Y. and Zhang, C. and Zhang, W. and Shao, C.-G.},
  journal = {Phys. Rev. D},
  volume  = {109},
  pages   = {064012},
  year    = {2024}
}

@article{RefQS27,
  author  = {Nozari, K. and Mehdipour, S. H.},
  journal = {Astrophys. Space Sci.},
  volume  = {339},
  pages   = {581},
  year    = {2012}
}

@article{RefQS28,
  author  = {Nicolini, P.},
  journal = {Int. J. Mod. Phys. A},
  volume  = {24},
  pages   = {1229},
  year    = {2019}
}

@article{RefQS29,
  author  = {Rayimbaev, J. and Bokhari, A. H. and Ahmedov, B.},
  journal = {Class. Quantum Gravity},
  volume  = {39},
  pages   = {075021},
  year    = {2022}
}

@article{metric,
  author  = {Charmousis, C. and Crisostomi, M. and Gregory, R. and Stergioulas, N.},
  journal = {Phys. Rev. D},
  volume  = {100},
  number  = {8},
  pages   = {084020},
  year    = {2019}
}

@article{B1,
  author  = {Ashraf, A. and Bouzenada, A. and Maurya, S. K. and Atamurotov, F. and Channuie, P. and Abd-Elmonem, A. and Abdalla, N. S. E.},
  journal = {Phys. Dark Universe},
  volume  = {47},
  pages   = {101787},
  year    = {2025}
}

@article{B12,
  author  = {Ditta, A. and others},
  journal = {Nucl. Phys. B},
  volume  = {1018},
  pages   = {117059},
  year    = {2025}
}
\bibliographystyle{ieeetr}

\end{document}